\documentclass[12pt]{article}

\usepackage{framed}
\usepackage{algorithm}
\usepackage[noend]{algpseudocode}
\usepackage{gensymb}

\makeatletter
\def\BState{\State\hskip-\ALG@thistlm}
\makeatother
\makeatletter
\newcounter{phase}[algorithm]
\newlength{\phaserulewidth}
\newcommand{\setphaserulewidth}{\setlength{\phaserulewidth}}

\makeatother
\setphaserulewidth{.7pt}

\newcounter{case}[algorithm]
\newlength{\caserulewidth}
\newcommand{\setcaserulewidth}{\setlength{\caserulewidth}}

\makeatother
\setcaserulewidth{.7pt}

\usepackage[english]{babel}
\usepackage[utf8x]{inputenc}
\usepackage[T1]{fontenc}
\usepackage{apacite}
\usepackage{bbding}
\usepackage{mathrsfs}
\usepackage{listings}
\usepackage{color}
\usepackage{bbm}
\usepackage{adjustbox}
\usepackage{epsfig, epsf, graphicx}
\usepackage{tikz}
\usepackage{animate}
\usepackage{pstricks, pst-node, psfrag}
\usepackage{amssymb,amsmath,bm}
\usepackage{amsthm}
\usepackage{verbatim,enumerate}
\usepackage{rotating, lscape}
\usepackage{diagbox}
\usepackage[doublespacing]{setspace}
\usepackage{tikz}

\usetikzlibrary{arrows,calc,tikzmark}
\usepackage[colorlinks=true,linkcolor=blue,citecolor=blue]{hyperref}%
\usepackage{float}
\usepackage[small]{caption}
\usepackage{subcaption}
\usepackage[square]{natbib}
\usepackage{lipsum}
\usepackage{threeparttable}
\usepackage{multirow}
\usepackage[page]{appendix}
\newtheorem{definition}{Definition}
\newtheorem{theorem}{Theorem}
\newtheorem{lemma}{Lemma}

\def\diag{\hbox{diag}}

\def\diag{\hbox{diag}}

\def\boxit#1{\vbox{\hrule\hbox{\vrule\kern6pt
			\vbox{\kern6pt#1\kern6pt}\kern6pt\vrule}\hrule}}
\def\refhg{\hangindent=20pt\hangafter=1}

\def\refhg{\hangindent=20pt\hangafter=1}

\def\bse{\begin{eqnarray*}}
	\def\ese{\end{eqnarray*}}
\def\be{\begin{eqnarray}}
\def\ee{\end{eqnarray}}
\def\bq{\begin{equation}}
\def\eq{\end{equation}}
\def\bse{\begin{eqnarray*}}
	\def\ese{\end{eqnarray*}}

\renewcommand{\baselinestretch}{1.2} 

\begin{document}

\def\spacingset#1{\renewcommand{\baselinestretch}%
{#1}\small\normalsize} \spacingset{1}


\baselineskip=26pt 
\begin{center} {\Huge{\bf Global Depths for Irregularly Observed Multivariate Functional Data}}
\end{center}

\baselineskip=12pt 

\begin{center}\large
Zhuo Qu\footnote[1]{\baselineskip=10pt Statistics Program,
King Abdullah University of Science and Technology,
Thuwal 23955-6900, Saudi Arabia. 
E-mail: zhuo.qu@kaust.edu.sa, marc.genton@kaust.edu.sa\\
This research was supported by the
King Abdullah University of Science and Technology (KAUST).}, Wenlin Dai\footnote[2]{\baselineskip=10pt Institute of Statistics and Big Data, Renmin University of China, Beijing 100872, China.\\
E-mail: wenlin.dai@ruc.edu.cn } and Marc G.~Genton\textcolor{blue}{$^1$}
\end{center}
\baselineskip=20pt 
\vskip -3mm
\centerline{\today} 
\begin{center}
\vskip -2mm
{\large{\bf Abstract}}
\end{center}
\vskip -6mm
Two frameworks for multivariate functional depth based on multivariate depths are introduced in this paper. The first framework is multivariate functional integrated depth, and the second framework involves multivariate functional extremal depth, which is an extension of the extremal depth for univariate functional data. In each framework, global and local multivariate functional depths are proposed. The properties of population multivariate functional depths and consistency of finite sample depths to their population versions are established. In addition, finite sample depths under irregularly observed time grids are estimated. As a by-product, the simplified sparse functional boxplot and simplified intensity sparse functional boxplot are proposed for visualization without data reconstruction. A simulation study demonstrates the advantages of global multivariate functional depths over local multivariate functional depths in outlier detection and running time for big functional data. An application of our frameworks to cyclone tracks data demonstrates the excellent performance of our global multivariate functional depths.

\vspace{0.2cm}


\baselineskip=26pt
\pagenumbering{arabic}

\section{Introduction}
Functional data analysis (FDA, \citeauthor{ramsay2005functional} \citeyear{ramsay2005functional}) is a branch of statistics that analyzes data about curves, surfaces, or any object sampled over a continuum. In addition, if the domain of functional data extends from $\mathbb{R}$ to $\mathbb{R}^p~(p \geq 2)$, then the data object becomes a multivariate curve or surface. These data are then termed multivariate functional data (\citeauthor{jacques2014model} \citeyear{jacques2014model}). The main problems of interest in FDA include: 1) extracting the central tendency (\citeauthor{lopez2009concept} \citeyear{lopez2009concept}, \citeauthor{lopez2011half} \citeyear{lopez2011half}, \citeauthor{sun2011functional} \citeyear{sun2011functional}), 2) detecting possible outliers (\citeauthor{hubert2008outlier} \citeyear{hubert2008outlier}, \citeauthor{dai2018multivariate} \citeyear{dai2018multivariate}, \citeauthor{dai2019directional} \citeyear{dai2019directional}), 3) supervised and unsupervised learning (\citeauthor{jacques2013funclust} \citeyear{jacques2013funclust}, \citeauthor{qu2022robust} \citeyear{qu2022robust}), 4) functional linear and nonlinear models (\citeauthor{jacques2014model} \citeyear{jacques2014model}, \citeauthor{happ2018multivariate} \citeyear{happ2018multivariate}), and 5) time warping to capture time variation (\citeauthor{liu2009simultaneous} \citeyear{liu2009simultaneous}, \citeauthor{srivastava2011registration} \citeyear{srivastava2011registration}). Data ordering is an effective tool for solving problems 1)-3). However, ordering functional data is difficult because of the special data structure. The functional depth is a powerful nonparametric tool to tackle this challenge.

Univariate functional depths can be categorized into three types according to their definitions. First, they can be extended directly from statistical depths (\citeauthor{zuo2000performance} \citeyear{zuo2000performance}) for multivariate data; examples of this type are L$^\infty$ depth (L$^{\infty}$D, \citeauthor{zuo2000general} \citeyear{zuo2000general}, \citeauthor{long2015study} \citeyear{long2015study}), h-mode depth (h-MD, \citeauthor{cuevas2007robust} \citeyear{cuevas2007robust}), random Tukey depth (RTD, \citeauthor{cuesta2008random} \citeyear{cuesta2008random}), and functional spatial depth (FSD, \citeauthor{chakraborty2014data} \citeyear{chakraborty2014data}, \citeauthor{sguera2014spatial} \citeyear{sguera2014spatial}). Second, they can be extended from univariate ordering to the functional domain; examples of this type are integrated depth (ID, \citeauthor{fraiman2001trimmed} \citeyear{fraiman2001trimmed}), integrated dual depth (IDD, \citeauthor{cuevas2009depth} \citeyear{cuevas2009depth}), modified version of band depth (MBD, \citeauthor{lopez2009concept} \citeyear{lopez2009concept}) and half region depth (MHRD, \citeauthor{lopez2011half} \citeyear{lopez2011half}), extremal depth (ED, \citeauthor{narisetty2016extremal} \citeyear{narisetty2016extremal}), and total variation depth (TVD, \citeauthor{huang2019decomposition} \citeyear{huang2019decomposition}). Third, they can be proposed specially for functional data; examples of this type are band depth (BD, \citeauthor{lopez2009concept} \citeyear{lopez2009concept}) and half region depth (HRD, \citeauthor{lopez2011half} \citeyear{lopez2011half}). Other statistical depths for multivariate functional data can be characterized into two types. The first type is the pointwise depth (see \citeauthor{claeskens2014multivariate} \citeyear{claeskens2014multivariate},  \citeauthor{lopez2014simplicial} \citeyear{lopez2014simplicial}). Specifically, \citeauthor{claeskens2014multivariate} (\citeyear{claeskens2014multivariate}) proposed a general framework for building multivariate functional depth (MFD) from pointwise multivariate depths, where MFD is the weighted integral of multivariate depths. The second type is the componentwise depth. For example, the multivariate band depth measure (\citeauthor{ieva2013depth} \citeyear{ieva2013depth}) is the average of the depth measures of marginal components.

When (univariate or) multivariate functional depths are extended from the (univariate or) multivariate depth, observations per subject are transformed into an empirical distribution of pointwise depths. Then, the notions of univariate (or multivariate) functional depth can be considered as a scalar extracted from the empirical distribution of pointwise depths. Generally, such methods can be categorized into integral and nonintegral methods. To integrate the functions of pointwise depths at each time point, ID, IDD, BD, MBD, and TVD are employed in the case of univariate functional depths, and MSBD and MFHD in the multivariate functional depths. With the nonintegral method, concepts of depth are introduced from diverse perspectives in univariate functional cases. For example, if we apply the minimum between the hypograph and epigraph (\citeauthor{lopez2011half} \citeyear{lopez2011half}), then we derive HRD (MHRD). Assuming that the functional depth of a curve $\bm{Y}$ is inversely proportional to the expectation of the supremum norm of the difference between $\bm{Y}$ and the finite samples from the same distribution, we obtain the L$^\infty$ depth (\citeauthor{long2015study} \citeyear{long2015study}). If we prioritize extreme values and rank the univariate functional data by the criterion based on the depth cumulative distribution functions  (cdf, \citeauthor{narisetty2016extremal} \citeyear{narisetty2016extremal}), then we obtain the rank-based extremal depth.

\begin{figure}
    \centering
    \vspace{-1cm}
    \includegraphics[width = 0.7\linewidth]{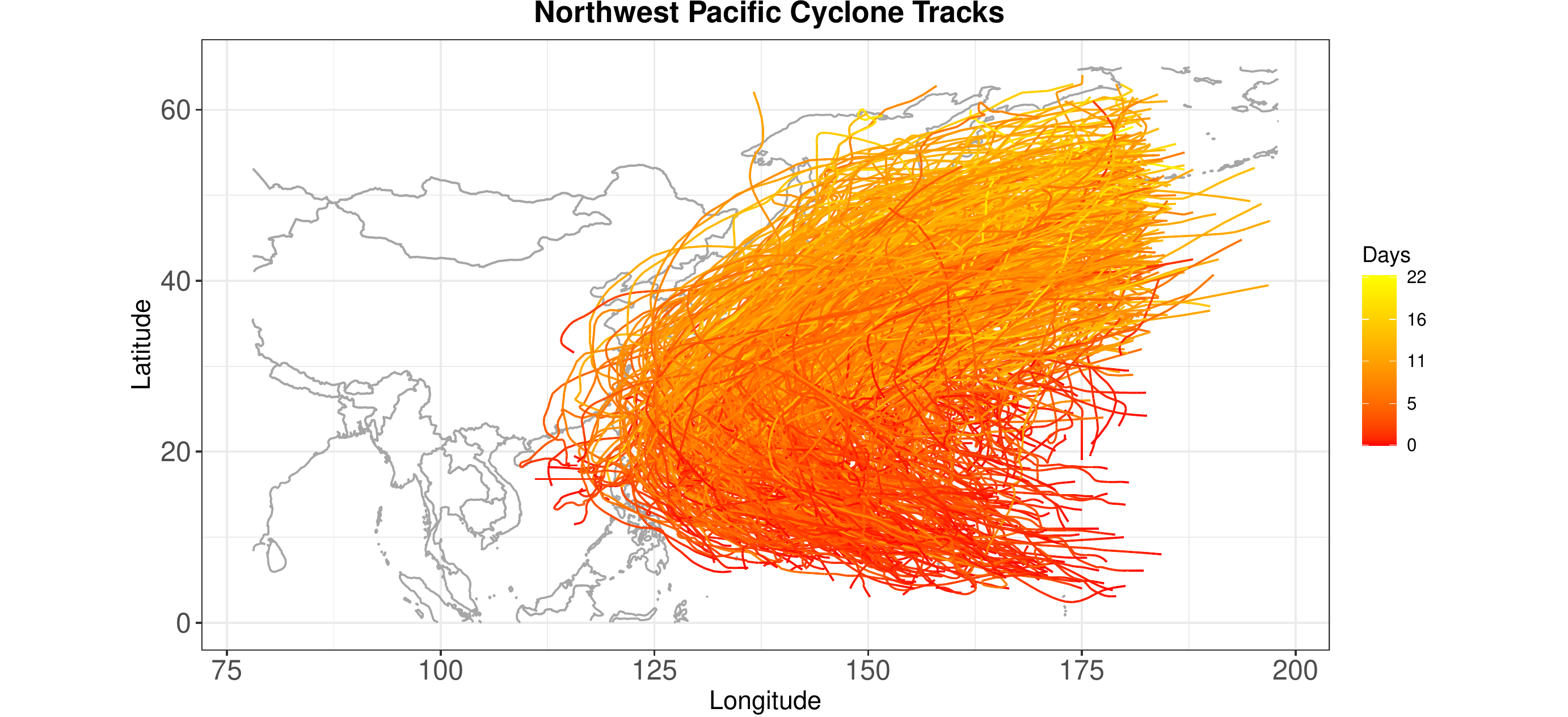}
    \caption{Northwest Pacific cyclone tracks indexed by days. The color starts with red and turns yellow as the sampling index increases.}
    \label{clusters}
\end{figure}

Certain issues demand our attention. First, the existing framework of building multivariate functional depth is mainly based on the integral of pointwise multivariate depths. Second, there are substantial restrictions to the implementation of functional depths. Indeed, usually functional depth notions require data to be observed on common time grids. However, this requirement often cannot be met in actual applications; see, for instance, the pattern of Northwest Pacific cyclone tracks (\citeauthor{qu2022robust} \citeyear{qu2022robust}) in Figure \ref{clusters}. Suggested solutions for this dilemma are the use of the revised depth for sparse multivariate functional data (\citeauthor{qu2022sparse} \citeyear{qu2022sparse}) and the integrated depth for partially observed functional data (POIFD, \citeauthor{elias2022integrated} \citeyear{elias2022integrated}). However, data reconstruction may cause loss to the precise description of outlier information, and POIFD requires that the partially observed functional data are evaluated on common time grids.

Two ideas can be explored when dealing with the depth of irregularly observed multivariate functional data. The first is whether there are frameworks other than the integrated depth, and the second is whether there exists an alternative that does not require data reconstruction. The solutions to the above problems are the important contributions of this study. The aim of this paper is to propose two frameworks of multivariate functional depths and to identify equivalent formulas under each framework. Taking the integrated depth in multivariate functional data as an example, the equivalent formulas of the global MFID (GMFID) and local MFID (LMFID) can be obtained. The typical properties of these depth notions and their consistency are examined. In addition, the simplified sparse functional boxplot and simplified intensity sparse functional boxplot are obtained based on the proposed depths for the purpose of visualization.

The rest of the paper is organized as follows. Section \ref{section2} proposes two formulas of multivariate functional depths in the population and finite sample versions, with proofs of the related theoretical properties in the Supplementary Material. Section \ref{section3} presents a simulation study where several depth functions are compared. Section \ref{section4} introduces extensions of the sparse functional boxplot and the intensity sparse functional boxplot without the requirement for data reconstruction. Section \ref{section5} illustrates the performance of the aforementioned new depths with an application to cyclone tracks data. Section \ref{section6} provides the conclusion and possible extensions of the current research.

\section{\large Definitions and Properties of Multivariate Functional Depths}
\label{section2}
\subsection{Notation}
\label{pointwise}
Denote the multivariate functional data as ${\bm{\mathcal{Y}}}(t)=(\mathcal{Y}^{(1)}(t),\ldots, \mathcal{Y}^{(p)}(t))^\top \in \mathbb{R}^p~(p \geq 1)$, $t \in \mathcal{T}$, where $\mathcal{T}$ is a compact set in $\mathbb{R}^{d}$. We view ${\bm{\mathcal{Y}}}(t): \mathcal{T} \xrightarrow[]{}\mathbb{R}^p$ as a realization of a $p$-variate random variable for $t \in \mathcal{T}$ from the associated distribution $F_{\bm{\mathcal{Y}}(t)}$. Assume that $\mathcal{Y}^{(i)}(t)~(i = 1, \ldots, p)$ is square integrable in $\mathcal{T}$, i.e., $\mathcal{Y}^{(i)}(t) \in L^2(\mathcal{T})$, and $\mathcal{H}$ is a Hilbert space defined as $\mathcal{H}:= L^2(\mathcal{T})\times \cdots \times L^2(\mathcal{T})$. We have $\bm{\mathcal{Y}}=\{\bm{\mathcal{Y}}(t)\}_{t \in \mathcal{T}}$ on $\mathcal{H}$ from the cumulative distribution function (cdf) $F_{\bm{\mathcal{Y}}}$. Under the integrability condition, the mean and covariance functions in a pointwise manner at $s, t~(s, t \in \mathcal{T})$ are defined as ${\bm{\mu}}(t) = {\rm E}\{\bm{\mathcal{Y}}(t)\}$ and $C_{ij}(s, t) = {\rm cov}\{\mathcal{Y}^{(i)}(s), \mathcal{Y}^{(j)}(t)\}$, respectively. Because $t \in \mathcal{T}$, $\mathcal{Y}^{(i)}$ is said to be continuous in mean-square at $t$ if $\lim\limits_{s\to t}{\rm E} \{{\big |}\mathcal{Y}^{(i)}(s)-\mathcal{Y}^{(i)}(t){\big |}^{2}\}=0$. Then, $\bm{\mu}$ is continuous, and the covariance function defines a covariance operator ${\bm{\mathcal {C}}}: \mathcal{H}\to \mathcal{H}$ for $\bm{f} \in \mathcal{H}$: $(\bm{{\mathcal {C}}}\bm{f})^{(j)}(t)=\sum\limits_{i=1}^p\int_{\mathcal{T}}C_{ij}(s,t)f^{(i)}(s)\,\mathrm {d} s$. The spectral theorem (Theorem 2.6 in \citeauthor{bachman2000functional} \citeyear{bachman2000functional}) applies to $\bm{\mathcal{C}}$, yielding eigenpairs $(\nu_{j},\bm{\varphi}_{j})$, where $\bm{\varphi}_{j}=(\varphi^{(1)}_j, \ldots, \varphi^{(p)}_j)^\top \in \mathcal{H}$ are eigenfunctions, which correspond to the nonnegative eigenvalues of $\bm{\mathcal{C}}$, $\nu_j$, in a nonincreasing order. Hence, we write $\displaystyle {C_{jj}(s, t)}=\sum _{m=1}^{\infty }\nu_j \varphi^{(j)}_{m}(s) \varphi^{(j)}_{m}(t)$. 

The multivariate Karhunen$-$Lo{\`e}ve decomposition (\citeauthor{happ2018multivariate} \citeyear{happ2018multivariate}) can be applied based on the above conditions. In addition, we incorporate uncorrelated multivariate measurement errors with mean $\bm{0}$ and variance matrix $\bm{V} = \diag\{\sigma_1^2, \ldots, \sigma^2_p\}$ to reflect additive measurement errors. Assuming that $\bm{Y}(t)$ has the distribution $F_{\bm{\mathcal{Y}}}(t)$ for $t \in \mathcal{T}$, we can denote it as 
\begin{equation}
\bm{Y}(t)=\bm{\mu}(t) +\sum_{m=1}^{\infty} \rho_{m} \bm{\varphi}_{m}(t) + \bm{\epsilon}(t),
\label{decomposition_y}
\end{equation}
where multivariate functional principal component scores $\rho_{m}$ serve as the weights of $\bm{\varphi}_m$.

In addition, we consider $\mathcal{C}(\mathcal{T})$ as the set of continuous functions on $\mathcal{T}$ and $\mathcal{C}^p(\mathcal{T})$ as the $p$-vectors of continuous functions on $\mathcal{T}$. In the population, we assume that $\bm{Y}$ is generated from the stochastic process $\bm{\mathcal{Y}}$ on $\mathcal{C}^p({\mathcal{T}})$ with the cdf $F_{\bm{\mathcal{Y}}}$. In the finite sample version, we assume that the $i$th ($i=1, \ldots, N$) sample $\bm{Y}_{i}(t_{i,k})$~($k=1, \ldots, T_i$) belongs to the distribution of $\bm{\mathcal{Y}}(t_{i,k})$ denoted by $F_{\bm{\mathcal{Y}}(t_{i,k}), N}$, where $T_i$ denotes the number of observations for sample $\bm{Y}_i$.

\subsection{Population Definitions}
A statistical depth function enables ordering from the center outwards and therefore the most central object has the largest depth and the least central object has the smallest depth. We list the condition of the statistical depth function as Condition (A). In addition, the assumption of the time density function in Condition (B.1) and the assumption of probability of $\bm{\mathcal{Y}}(t)$ in Condition (C) are provided in the following paragraphs before defining the multivariate functional depths.

\textbf{Condition (A)}. Let $D(\cdot; F_{\bm{\mathcal{X}}}): \mathbb{R}^p → [0, 1]$ be a statistical depth function for the probability distribution of a $p$-variate random vector $\mathcal{X}$ with cdf $F_{\bm{\mathcal{X}}}$ following \citeauthor{zuo2000general} (\citeyear{zuo2000general}). Assume that the depth function $D$ satisfies the following properties: affine invariance, maximality at the center, monotonicity relative to the deepest point, and vanishing at infinity. Further, for the depth function $D$, there exists the depth region $D_\beta(F_{\bm{\mathcal{X}}})$ (Definition 3.1 in \citeauthor{zuo2000structural} \citeyear{zuo2000structural}) at level $\beta >0$, defined as $D_{\beta}(F_{\bm{\mathcal{X}}}) = \{\bm{x} \in \mathbb{R}^p: D(\bm{x}; F_{\bm{\mathcal{X}}}) > \beta\}$.

\textbf{Condition (B.1)}. Let time points originate from a design density $g(u)$ such that $G(t)=\int_{-\infty}^{t}g(u){\rm d}u$. Let the compact set $\mathcal{T}=[G^{-1}(0), G^{-1}(1)]$.~We assume that $g$ is differentiable, $\inf\limits_{t \in \mathcal{T}} g(t)>0$, and $g$ possesses the Glivenko$-$Cantelli property (GC property, Theorem 19.1 in \citeauthor{van2000asymptotic} \citeyear{van2000asymptotic}) uniformly over convex sets.

\textbf{Condition (C)}. The probability $P_{\bm{\mathcal{Y}}(t)}(B)=P(\bm{\mathcal{Y}}(t) \in B)$ holds for any set $B$ from the Borel $\sigma$-algebra of $\mathbb{R}^p$. The probability $P_{\bm{\mathcal{Y}}}$ is absolutely continuous and $P_{\bm{\mathcal{Y}}(t)}$ has a unique deepest point for every $t \in \mathcal{T}$. Additionally, we assume that $P$ possesses the GC property uniformly over convex sets.

Based on Conditions (A), (B.1), and (C), we study two depth proposals for multivariate functional data: the multivariate functional integrated depth (MFID) and the multivariate functional extremal depth (MFED). MFID is based on the average pointwise depths, whereas MFED is based on the ranking criterion according to the cdfs of the pointwise depths per subject. Overall, depth is determined before ranking in MFID, whereas the ranking is used to obtain the depth in MFED.

First, we adopt the MFID proposed by \citeauthor{claeskens2014multivariate} (\citeyear{claeskens2014multivariate}) for regularly observed multivariate functional data (see \citeauthor{elias2022integrated} \citeyear{elias2022integrated} for partially observed multivariate functional data). Recall that the MFID of $\bm{X}$ in the population version is defined as \begin{equation}
MFID(\bm{X}; F_{\bm{\mathcal{Y}}}, \beta) = \int_{\mathcal{T}}D(\bm{X}(t); F_{\bm{\mathcal{Y}}(t)}) \cdot w_{\beta}(t; F_{\bm{\mathcal{Y}(t)}}){\rm d}t,
\label{MFID}
\end{equation}
where $w_{\beta}(t; F_{\bm{\mathcal{Y}}(t)})$ is a weight function satisfying $\int_{\mathcal{T}}w_{\beta}(t; F_{\bm{\mathcal{Y}}(t)}){\rm d}t = 1$ for a fixed $\beta \in (0, 1]$. Two common choices of $w_{\beta}(t; F_{\bm{\mathcal{Y}}(t)})$ are considered: the first type is proportional to the volume of the depth region (\citeauthor{claeskens2014multivariate} \citeyear{claeskens2014multivariate}) at time point $t$, and the second type is proportional to the time density (Condition (B.1)) at time point $t$. Hence,
\[w_{\beta}(t; F_{\bm{\mathcal{Y}}(t)})=
\begin{cases}
\frac{{\rm vol}\{D_{\beta}(F_{\bm{\mathcal{Y}}(t)})\}}{\int_{\mathcal{T}}{\rm vol}\{D_{\beta}(F_{\bm{\mathcal{Y}}(u)})\}{\rm d}u}, & w \propto {\rm ~the~volume~of~the~depth~region~at~} t,\\
g(t), & w \propto {\rm ~time~density~at~} t.
\end{cases}
\]

Then, we use the depth cdf criterion from the extremal depth (\citeauthor{narisetty2016extremal} \citeyear{narisetty2016extremal}) in the univariate functional case. Instead of applying the criterion to pointwise univariate rank-based depths, we apply it to pointwise multivariate depths and propose multivariate functional extremal depth (MFED); see Definition \ref{MFD_population1}.

\begin{definition} \textbf{Multivariate Functional Extremal Depth} (MFED, Population Version). Assume that Conditions (A), (B.1), and (C) hold. Consider a $p$-variate stochastic process $\{\bm{\mathcal{Y}}(t)\}_{t \in \mathcal{T}}$ that generates continuous paths on $\mathcal{C}^p(\mathcal{T})$ with cdf $F_{\bm{\mathcal{Y}}}$. Let $D$ be a statistical depth function on $\mathbb{R}^p$. 
From an arbitrary $\bm{X} \in \mathcal{C}^p(\mathcal{T})$, the MFED of $\bm{X}$ is \vspace{-0.2cm}\begin{equation}
    MFED(\bm{X}; F_{\bm{\mathcal{Y}}})= 1- P(\bm{X}\prec \bm{Y})=P(\bm{X} \succeq \bm{Y}), {\rm~where~} \bm{Y} \sim F_{\bm{\mathcal{Y}}}.
    \label{GMFED}
\end{equation}
\label{MFD_population1}
\end{definition}
\vspace{-1.5cm}

We retain the notation $\prec$ as that used in \citeauthor{narisetty2016extremal} (\citeyear{narisetty2016extremal}), i.e., $\bm{X} \prec \bm{Y}$ means that $\bm{X}$ is more extreme than $\bm{Y}$; $\bm{X} \succeq \bm{Y}$ means that $\bm{X}$ is either equivalent to $\bm{Y}$ in terms of depth, i.e., $\bm{X} \sim \bm{Y}$, or $\bm{X}$ is less extreme than $\bm{Y}$, i.e., $\bm{X} \succ \bm{Y}$. The comparison is implemented through the depth cdfs $\Psi_{\bm{X}}$ of $D(\bm{X}; F_{\bm{\mathcal{Y}}})$ and $\Psi_{\bm{Y}}$ of $D(\bm{Y}; F_{\bm{\mathcal{Y}}})$, where $\Psi_{\bm{X}}(r)=\int_{\mathcal{T}} \bm{1}\{D(\bm{X}(t); F_{\bm{\mathcal{Y}}(t)}) \leq r\}{\rm d}t$, and $\bm{1}$ represents the indicator function. We list the criteria for comparing the extremal tendency between $\bm{X}$ and $\bm{Y}$ as in \citeauthor{narisetty2016extremal} (\citeyear{narisetty2016extremal}): 1) $\bm{X} \prec \bm{Y}$ denotes that there exists $0<r<1$ such that $\forall~0\leq u<r$, $\Psi_{\bm{X}}(u)=\Psi_{\bm{Y}}(u)$, and $\Psi_{\bm{X}}(r)>\Psi_{\bm{Y}}(r)$; 2) $\bm{X} \succ \bm{Y}$ denotes that there exists $0<r<1$ such that $\forall~0\leq u<r$, $\Psi_{\bm{X}}(u)=\Psi_{\bm{Y}}(u)$, and $\Psi_{\bm{X}}(r)<\Psi_{\bm{Y}}(r)$; and 3) $\bm{X} \sim \bm{Y}$ denotes that $\forall~ 0\leq u<1$, $\Psi_{\bm{X}}(u)=\Psi_{\bm{Y}}(u)$.

We observe that MFID and MFED are calculated based on the pointwise multivariate depths of the original data; that is, for each time index, we collect the data observed at such time index and then calculate the pointwise depths of such data. The multivariate depth calculation is implemented at each time index, if we do not normalize the data.

Consider normalizing $\bm{Y}(t)$ in a pointwise manner, i.e., 
\begin{equation}
    \bm{Y}^{*}(t) = \bm{\Sigma}_t^{-1/2}\{\bm{Y}(t) - \bm{\mu}(t)\}
    ,
    \label{standardization}
\end{equation}
where $\bm{\Sigma}_t = {\rm var}\{\bm{Y}(t)\} = \{C_{ij}(t, t)\}_{i,j = 1}^p + \bm{V}$ is symmetric and positive-definite. If we normalize the data, then the normalized data across different time indexes all satisfy the distribution of a $p$-variate random vector $\bm{\mathcal{S}}$ such that ${\textrm E}(\bm{\mathcal{S}})=\bm{0} \in \mathbb{R}^p$ and ${\rm var}(\bm{\mathcal{S}})=\textbf{I}_p$ with cdf $F_{\bm{\mathcal{S}}}$. Here, $\textbf{I}_p$ is the identity matrix with $p$ rows. The identical distribution of the normalized data may be beneficial for pointwise depth calculation. That is, we can combine all the normalized data across the time indexes and calculate their pointwise depths only once. The advantage is more apparent when data are irregularly observed. In that case, the pointwise depths of the original data might not be calculated accurately at some indexes either because of a lack of observations or because the observed samples are biased representations of the population. In extreme cases where there are less than $p+1$ observations at some index, multivariate depths cannot be calculated because there are too few observations.

Depending on the manner that the pointwise depths are obtained, we term the MFID and the MFED in Definition \ref{MFD_population1} as the local MFID (LMFID) and local MFED (LMFED), respectively. Further, we apply MFID and MFED to the normalized data and define new MFDs as the global MFID (GMFID) and global MFED (GMFED), respectively: 

\begin{definition} \textbf{Global Multivariate Functional Depth} (GMFD, Population Version). Assume that Conditions (A), (B.1), and (C) hold, and the stochastic process $\bm{\mathcal{Y}}^{*}(t)$ across time $t$ has cdf $F_{\bm{\mathcal{S}}}$. From an arbitrary $\bm{X} \in \mathcal{C}^p(\mathcal{T})$, we obtain $\bm{X}^*(t) \in \mathbb{R}^p ~\forall t \in \mathcal{T}$. 

1. The \textbf{global multivariate functional integrated depth} (GMFID) of $\bm{X}$ is
\begin{equation}
GMFID(\bm{X}; F_{\bm{\mathcal{Y}}}, \beta) = \int_{\mathcal{T}}D(\bm{X}^{*}(t); F_{\bm{\mathcal{S}}}) \cdot w_{\beta}(t; F_{\bm{\mathcal{S}}}){\rm d}t,
\label{GMFID}
\end{equation}
where $w_{\beta}(t; F_{\bm{\mathcal{S}}})$ is the definition of $w_{\beta}(t; F_{\bm{\mathcal{Y}}(t)})$ where cdf $F_{\bm{\mathcal{S}}}$ replaces $F_{\bm{\mathcal{Y}}(t)}$.

2. The \textbf{global multivariate functional extremal depth} (GMFED) of $\bm{X}$ is defined in the same way as in Eq. (\ref{GMFED}). The only difference is that we replace the depth cdf $\Psi_{\bm{X}}$ with $\Psi_{\bm{X}^*}$, where $\Psi_{\bm{X}^*}(r)=\int_{\mathcal{T}} \bm{1}\{D(\bm{X}^*(t); F_{\bm{\mathcal{S}}}) \leq r\}{\rm d}t$.
\label{MFD_population2}
\end{definition}
Clearly, GMFID is equivalent to LMFID when $w$ is proportional to the time density at $t$, and GMFED is always equivalent to LMFED. When $w$ is proportional to the volume of the depth region at $t$, usually GMFID is not equivalent to LMFID unless the data are homoscedastic across the time indexes. The discussion about the equivalence is based on the affine invariance of $D$ from Condition (A). 

First, the affine invariance property of $D$ leads to $D(\bm{X}(t); F_{\bm{\mathcal{Y}}})=D(\bm{X}^{*}(t); F_{\bm{\mathcal{S}}})$. If $w$ is only proportional to the time density at $t$, then the normalization of data does not cause a change in $w$, and in this case, GMFID is equivalent to LMFID. In addition, $D(\bm{X}(t); F_{\bm{\mathcal{Y}}})=D(\bm{X}^{*}(t); F_{\bm{\mathcal{S}}})$ results in $\Psi_{\bm{X}}(r)=\Psi_{\bm{X}^*}(r)$. Hence, GMFED is always equivalent to LMFED. Second, $\textrm{vol}\{D_{\beta}(F_{\bm{\mathcal{Y}}(t)})\}={\rm det}|\bm{A}(t)|\cdot \textrm{vol}\{D_{\beta}(F_{\bm{A}(t)\bm{\mathcal{Y}}(t)+\bm{B}(t)})\}$, with $\bm{A}(t)=\bm{\Sigma}^{-1/2}_t$ and $\bm{B}(t)=-\bm{\Sigma}^{-1/2}_t \bm{\mu}(t)$. When $w$ is proportional to the volume of the depth region at $t$, the equivalence between $w_{\beta}(t; F_{\bm{\mathcal{Y}}(t)})$ and $w_{\beta}(t; F_{\bm{\mathcal{S}}})$ is not guaranteed unless the data are homoscedastic regardless of their time indexes.

The depth properties of LMFID when $w$ is proportional to the volume of the depth region have been studied; see Theorem \ref{depth in population} in \citeauthor{claeskens2014multivariate} (\citeyear{claeskens2014multivariate}). Hence, we only need to analyze the depth properties of the remaining five depths by considering the definitions of $w$ in the integrated depth. Because of the equivalence between two pairs (GMFED, LMFED) and (GMFID, LMFID) when $w$ is proportional to the time density at $t$, examining the properties of global MFDs (GMFDs) is sufficient. The depth properties of GMFID and GMFED are demonstrated in Theorem \ref{depth in population}. 
 
\begin{theorem} 
\label{depth in population}
Assume that Conditions (A), (B.1), and (C) hold. Then GMFID and GMFED, as defined in Definition \ref{MFD_population2}, are statistical depth functions (GMFDs) satisfying the following key properties:
 ~\\
 1) \textbf{Affine invariance}. $GMFD(\bm{X}; F_{\bm{\mathcal{Y}}}, \beta)=GMFD(\bm{A}\bm{X}+\widetilde{\bm{X}}; F_{\bm{A}\bm{\mathcal{Y}}+\widetilde{\bm{\mathcal{Y}}}}, \beta)$ with the stochastic process $\{\bm{A}\bm{\mathcal{Y}}(t)+\widetilde{\bm{\mathcal{Y}}}(t), t \in \mathcal{T}\}$ on $\mathcal{C}^p(\mathcal{T})$ from cdf $F_{\bm{A}\bm{\mathcal{Y}}+\widetilde{\bm{\mathcal{Y}}}}$, and any matrix $\bm{A} \in \mathbb{R}^{p \times p}$ with ${\rm det}(\bm{A}) \neq 0$.
 ~\\
 2) \textbf{Maximality at the center}. $GMFD(\bm{\Theta}; F_{\bm{\mathcal{Y}}}, \beta) = \sup\limits_{\bm{X} \in \mathcal{C}^p(\mathcal{T})} GMFD(\bm{X}; F_{\bm{\mathcal{Y}}}, \beta)$, for any distribution $F_{\bm{\mathcal{Y}}}$ that has a uniquely defined point of symmetry $\bm{\Theta}$.
 ~\\
 3) \textbf{Monotonicity relative to the deepest point}. $GMFD(\bm{X}; F_{\bm{\mathcal{Y}}}, \beta) \leq GMFD(\bm{\Theta} + a(\bm{X} - \bm{\Theta}); F_{\bm{\mathcal{Y}}}, \beta)$, for any distribution $F_{\bm{\mathcal{Y}}}$ with a deepest point $\bm{\Theta}$ and for any $a \in [0, 1]$.
 ~\\
 4) \textbf{Vanishing at infinity}. For $1 \leq k \leq p$ and for a series of curves $\{X^{(k)}_{n}\}$ with $\lim\limits_{n \to \infty}|X^{(k)}_{n}(t)| = \infty$ for almost all time points $t \in \mathcal{T}$: $\lim\limits_{n\to \infty} GMFD(\bm{X}_{n}; F_{\bm{\mathcal{Y}}}, \beta) = 0$. 
 \end{theorem}

Theorem \ref{median} states that there exists a deepest GMFD curve at absolutely continuous distributions $F_{\bm{\mathcal{Y}}}$ with a unique deepest point at each time point. Naturally, $\bm{\Theta}$ in Theorem~\ref{theorem_center} is defined as the median of $\bm{\mathcal{Y}}$, with the maximum value of $D(\cdot; F_{\bm{\mathcal{Y}}(t)})$ at each time point~$t$.

\begin{theorem} \label{median}
Assume that Conditions (A), (B.1), and (C) hold. Consider the curve $\bm{\Theta}$ that equals the vector in $\mathbb{R}^p$ with the maximum value of $D(\cdot; F_{\bm{\mathcal{Y}}(t)})$ at each time point $t$. Then, the following hold:

1. If the process $\bm{\mathcal{Y}}$ is such that $\mathcal{T}\times \mathbb{R}^p \to \mathbb{R}:(t, \bm{x})\mapsto D(\bm{x}; F_{\bm{\mathcal{Y}}(t)})$ is continuous, then $\bm{\Theta}$ is continuous: $\bm{\Theta} \in \mathcal{C}^p(\mathcal{T})$.

2. $\bm{\Theta}$ has maximal $GMFD$: for all $\bm{X} \in \mathcal{C}^p(\mathcal{T})$, $GMFD(\bm{X}; F_{\bm{\mathcal{Y}}}, \beta) \leq GMFD(\bm{\Theta}; F_{\bm{\mathcal{Y}}}, \beta)$.

3. Any curve $\widetilde{\bm{\Theta}} \in \mathcal{C}^p(\mathcal{T})$ with maximal GMFD should have maximal D at each time point $t$, i.e., $D(\widetilde{\bm{\Theta}}(t); F_{\bm{\mathcal{Y}}(t)})=\max\limits_{\bm{x} \in \mathbb{R}^p}D(\bm{x}; F_{\bm{\mathcal{Y}}(t)})$. 
\label{theorem_center}
\end{theorem}

The following subsection proposes the finite sample definitions of LMFDs and GMFDs and provides the consistency of the finite sample depths to the population depths.

\subsection{Finite Sample Definitions}
\label{sample_version}
Let $S:=\{\bm{Y}_i\}_{i=1}^N$ be a set of $N$ observations with cdf $F_{\bm{\mathcal{Y}}(t), N}$ at each time point $t$. In practice, one usually does not have identical time grids for each observation. Let $\bm{Y}_i(t_{i,k})$ be the $k$th~($k=1, \ldots, T_i$) observation of the random function $\bm{Y}_i$, made at a random time $t_{i,k} \in \mathcal{T}$. The observed time points for the $i$th random curve are $\bm{t}_i=\{t_{i, k}\}_{k=1}^{T_i}$. Let  $T_s=\min\limits_{i \in \{1,\ldots, N\}}T_i$, $\overline{T}=\frac{\sum_{i=1}^N T_i}{N}$ , $t^N_1=\min\limits_{\substack{k = 1, \ldots, T_i\\i=1, \ldots, N}} t_{i,k}$, and $t^{N}_{\overline{T}}=\max\limits_{\substack{k = 1, \ldots, T_i\\i=1, \ldots, N}} t_{i,k}$. Then, the compact set $\mathcal{T}^N$ is defined as $\mathcal{T}^N=[t^{N}_{1}, t^N_{\overline{T}}]$, and let $t^{N}_{0}=t^{N}_{1}$ and $t^N_{\overline{T}+1}=t^N_{\overline{T}}$.

The $i$th ($i=1, \ldots, N$) random curve we consider is \begin{equation}
\bm{Y}_{i}(t_{i,k})=\bm{\mu}(t_{i,k}) +\sum_{m=1}^{\infty} \rho_{i,m} \bm{\varphi}_{m}(t_{i,k}) + \bm{\epsilon}_{i,k}, ~~k = 1,\ldots, T_i,~m \in \mathbb{Z}_{+},
\end{equation}
where $\bm{\varphi}_{m}(t_{i,k})$ is the $m$th eigenfunction evaluated at $t_{i,k}$, and $\bm{\epsilon}_{i,k}$ are the additional measurement errors that are assumed to be i.i.d. and independent of $\rho_{i, m}$. 

Several conditions are listed for local multivariate functional depth definitions.

\textbf{Condition (B.2)}. In the finite sample version, let the time points originate from a design function $g(t)$ in Condition (B.1). Then, $G(t)=\int_{-\infty}^{t}g(u){\rm d}u$ is satisfied such that $t^N_j=G^{-1}(j/\overline{T})$, and $\mathcal{T}^N=[t^N_1, t^N_{\overline{T}}]$. When $T_i \to \infty$ $\forall~i=1, \ldots, N$, $\overline{T} \to \infty$. In addition, $t^N_1$ converges to $G^{-1}(0)$ from the right side, and $t^N_{\overline{T}}$ converges to $G^{-1}(1)$ from the left side such that $\mathcal{T}^N \subseteq \mathcal{T}$ where $\mathcal{T}=[G^{-1}(0), G^{-1}(1)]$. 
 
\textbf{Condition (D)}. The statistical depth $D$ satisfies the four properties in Condition (A), $\sup\limits_{\bm{x} \in \mathbb{R}^p}|D(\bm{x}; F_{N})-D(\bm{x}; F)| \xrightarrow[]{a.s.} 0$ for the cdf $F_{N} \xrightarrow[]{} F$ as $N \to \infty$, and $P(\{\bm{x} \in \mathbb{R}^p: D(\bm{x}, F)= \beta\})=0$ for any fixed $\beta \in [0, 1]$.

\textbf{Condition (E)}. Let $cs$ be the total number of functional crossings by any pair of functions,
where $N$ is the number of sample functions. We assume that $cs = \exp\{O_p (N)\}$.

The sample LMFID evaluated on identical time grids when $w$ is proportional to the volume of the depth region at $t$ was proposed by \citeauthor{claeskens2014multivariate} (\citeyear{claeskens2014multivariate}). To make this sample LMFID more general for our irregular time grids and to consider the LMFED in the finite sample version, we propose the following sample LMFDs in Definition \ref{finite_LMFD}.

\begin{definition} \textbf{Local Multivariate Functional Depth} (LMFD$_N$, Finite Sample Version). For a sample of multivariate curve observations $\{\bm{Y}_{i}(t);i = 1, \ldots, N\}_{t \in \mathcal{T}^N}$ with cdf $F_{\bm{\mathcal{Y}}(t),N}$ at each time point $t$, the following assumptions are made. 

1. Assume that Conditions (A), (B.1), (B.2), (C), and (D) hold. The \textbf{sample local multivariate functional integrated depth} (LMFID$_N$) at $\bm{X} \in \mathcal{C}^p(\mathcal{T})$ is 
\begin{equation}
LMFID_N(\bm{X}; F_{\bm{\mathcal{Y}}, N}, \beta) = \sum_{j=1}^{\overline{T}}D(\bm{X}(t^N_{j}); F_{\bm{\mathcal{Y}}(t^N_{j}),N})w_{\beta}(t^N_{j}; F_{\bm{\mathcal{Y}}(t^N_j), N}),
\end{equation}
where \[w_{\beta}(t^{N}_{j}; F_{\bm{\mathcal{Y}}(t^N_{j}),N})=
\begin{cases}
\frac{{\rm vol}\{D_{\beta}(F_{\bm{\mathcal{Y}}(t^N_{j}),N})\}(t^N_{j+1} − t^N_{j-1})}{\sum_{j=1}^{\overline{T}}{\rm vol}\{D_{\beta}(F_{\bm{\mathcal{Y}}(t^N_{j}),N})\}(t^N_{j+1} − t^N_{j−1})}, & w \propto {\rm vol}\{D_{\beta}(F_{\bm{\mathcal{Y}}(t^N_{j}),N})\},\\
\frac{g_{N,j}(t^N_j)(t^N_{j+1}-t^N_j)}{\sum_{j=1}^{\overline{T}}g_{N,j}(t^N_j)(t^N_{j+1}-t^N_j)}, & w \propto {\rm ~} g_{N,j}(t^N_j),
\end{cases}
\]
$g_{N,j}(t)=\sum_{e=1}^N \sum_{h=1}^{T_e} \bm{1}(t_{e,h}\in \mathcal{T}^{N}_{k})$, and $\mathcal{T}^{N}_{k}:=[{t}^N_{k-1}, {t}^N_{k}]$.

2. Assume that Conditions (A), (B.1), (B.2), (C), and (E) hold. Let $\Psi_{\bm{X}}(r)=\int_{\mathcal{T}} \bm{1}\{D(\bm{X}(t); F_{\bm{\mathcal{Y}}(t), N}) \leq r\}{\rm d}t$. The \textbf{sample local multivariate functional extremal depth} (LMFED$_N$) at $\bm{X} \in \mathcal{C}^p(\mathcal{T})$ is
\begin{equation}
    LMFED_N(\bm{X}; F_{\bm{\mathcal{Y}}, N})= 1- \frac{\sum_{i=1}^N\bm{1}(\bm{X} 	\prec \bm{Y}_i)}{N}=\frac{\sum_{i=1}^N\bm{1}(\bm{X} 	\succeq \bm{Y}_i)}{N},~ \bm{Y}_i \sim F_{\bm{\mathcal{Y}}, N}.
    \label{LMFED}
\end{equation}
\label{finite_LMFD}
\end{definition}
\vspace{-1cm}
The sample GMFDs are defined after normalizing $\bm{Y}_{i}(t_{i,k})$.~One notable difference is that the sample mean and sample covariance at each time are unknown and should be estimated. The pointwise normalization for $\bm{Y}_{i}(t_{i,k})$ with $i=1,\ldots,N$ and $k = 1, \ldots, T_i$ is
\begin{equation}
    \widehat{\bm{Y}}^{*}_{i}(t_{i,k}) = \bm{Q}_{\mathcal{T}^N_{j}}^{-1/2}\{\bm{Y}(t_{i,k}) - \overline{\bm{Y}}(\mathcal{T}^N_{j})\}, ~~t_{i,k} \in \mathcal{T}^N_j.
    \label{standardization_sample}
\end{equation}
Here, $\overline{Y^{(l)}}(\mathcal{T}^N_{j})=\frac{\sum_{g=1}^N \sum_{h=1}^{T_g} Y^{(l)}_g(t_{g,h})\bm{1}(t_{g,h}\in \mathcal{T}^{N}_{j})}{\sum_{g=1}^N \sum_{h=1}^{T_g} \bm{1}(t_{g,h}\in \mathcal{T}^{N}_{j})}$ for $l=1,\ldots, p$ and the sample mean $\overline{\bm{Y}}(\mathcal{T}^N_{j})=(\overline{Y^{(1)}}(\mathcal{T}^N_{j}), \ldots, \overline{Y^{(p)}}(\mathcal{T}^N_{j}))^\top$, where there exists a unique $j$ from $1, \ldots, \overline{T}$ such that $t_{i,k}\in \mathcal{T}^{N}_{j}$.~The pointwise sample covariance $\bm{Q}_{\mathcal{T}^N_{j}} \in \mathbb{R}^{p\times p}$, and its $(l,m)$th element ($l,m=1,\ldots,p$) is $\overline{Y^{(l)}Y^{(m)}}(\mathcal{T}^{N}_{j})-\overline{Y^{(l)}}(\mathcal{T}^{N}_{j})\cdot\overline{Y^{(m)}}(\mathcal{T}^{N}_{j})$, where the quantity $\overline{Y^{(l)}Y^{(m)}}(\mathcal{T}^N_{j})=\frac{\sum_{g=1}^N \sum_{h=1}^{T_g} Y^{(l)}_{g}(t_{g,h})Y^{(m)}_{g}(t_{g,h})\bm{1}(t_{g,h}\in \mathcal{T}^{N}_{j})}{\sum_{g=1}^N \sum_{h=1}^{T_g} \bm{1}(t_{g,h}\in \mathcal{T}^{N}_{j})}$. To estimate the sample mean and covariance at any time $t$, we rely on the observations in the interval $\mathcal{T}^{N}_{j}$ that includes $t$. 

The stochastic process $\{\widehat{\bm{\mathcal{Y}}}^{*}_{i}(t_{i,k})\}_{t_{i,k} \in \mathcal{T}^N}$ at each time point $t_{i,k}$ has cdf $F_{\widehat{\bm{\mathcal{Y}}}^*(t_{i,k}), N}$. In Lemma \ref{consistency_measure}, we prove that the sample mean and sample covariance converge almost surely to the population sample and population covariance.

\begin{lemma} (Consistency of the sample mean and covariance). For $t \in \mathcal{T}^N$, there exists a unique $\mathcal{T}^N_j$ such that $t \in \mathcal{T}^N_j$. Assuming that Condition (B.2) holds, we have the following:

1) For $l = 1, \ldots, p$, $\overline{Y^{(l)}}(\mathcal{T}^N_{j}) \xrightarrow[]{a.s.} {\rm E}\{Y^{(l)}(t)\}$ as $T_s \to \infty$ and $N \to \infty$;

2) $\bm{Q}_{\mathcal{T}^N_{j}} \in \mathbb{R}^{p \times p} \xrightarrow[]{a.s.} {\rm var}\{\bm{Y}(t)\}$ as $T_s \to \infty$ and $N \to \infty$, i.e. $\forall 1\leq l,m \leq p$, $\overline{Y^{(l)}Y^{(m)}}(\mathcal{T}^N_{j})-\overline{Y^{(l)}}(\mathcal{T}^N_{j})\cdot \overline{Y^{(m)}}(\mathcal{T}^N_{j}) \xrightarrow[]{a.s.} {\rm cov}\{Y^{(l)}(t), Y^{(m)}(t)\}$.
\label{consistency_measure}
\end{lemma}

Lemma \ref{consistency_measure} infers the convergence of the normalized data in a small bin to a $p$-variate random vector with zero mean and unit covariance in distribution when the length of the small bin converges to $0$ and the number of samples tends to infinity. Hence, we first propose the sample GMFDs with the pointwise normalized data following the distribution at the corresponding index; see Definition \ref{finite_GMFD}. Subsequently, we prove the consistency of the sample GMFDs to the population GMFDs. 
\begin{definition} \textbf{Global Multivariate Functional Depth} (GMFD$_N$, Finite Sample Version). For a sample of multivariate curves $\{\widehat{\bm{Y}}^{*}_i(t); i = 1, \ldots, N\}_{t \in \mathcal{T}^N}$ with cdf $F_{\widetilde{\bm{\mathcal{Y}}}^{*}(t),N}$ at each time point $t$. 

1. Assume that Conditions (A), (B.1), (B.2), (C), and (D) hold. The \textbf{sample global multivariate functional integrated depth}~(GMFID$_N$) at $\bm{X} \in \mathcal{C}^p(\mathcal{T})$ is 
\begin{equation}
GMFID_N(\bm{X}; F_{\bm{\mathcal{Y}}, N}, \beta) =\sum_{j=1}^{\overline{T}}D(\widehat{\bm{X}}^{*}(t^N_{j}); F_{\widehat{\bm{\mathcal{Y}}}^{*}(t^N_{j}),N})w_{\beta}(t^N_{j}; F_{\widehat{\bm{\mathcal{Y}}}^{*}(t^N_j), N}).
\end{equation}

2. Assume that Conditions (A), (B.1), (B.2), (C), and (E) hold. The \textbf{sample global multivariate functional extremal depth}~(GMFED$_N$) at $\bm{X} \in \mathcal{C}^p(\mathcal{T})$, $GMFED_N(\bm{X}; F_{\bm{\mathcal{Y}}, N})$, is defined as shown in Eq. (\ref{LMFED}). The only difference is that we apply the depth cdf for $\bm{X}^* \in \mathcal{C}^p(\mathcal{T})$ instead of $\bm{X}$, where $\Psi_{\bm{X}^*}(r)=\int_{\mathcal{T}} \bm{1}\{D(\bm{X}^*(t); F_{\bm{\mathcal{Y}}^*(t)}) \leq r\}{\rm d}t$.
\label{finite_GMFD}
\end{definition}

Again, for the finite sample version, GMFID$_N$ is equivalent to LMFID$_N$ when $w$ is proportional to the time density at $t$, and GMFED$_N$ is always equivalent to LMFED$_N$. Since the consistency of the finite LMFID$_N$ to its population version when $w$ is proportional to the volume of the depth region at $t$ was proved in Theorem~3 in \citeauthor{claeskens2014multivariate} (\citeyear{claeskens2014multivariate}), considering the consistency of the sample GMFD$_N$s to the corresponding population GMFDs under some conditions is sufficient; see Theorem \ref{consistency}.

 \begin{theorem} (Consistency of the sample depths). Let $\{\bm{Y}_i\}^N_{i=1}$ be a sample with the same distribution as $\bm{\mathcal{Y}} \in \mathcal{C}^p(\mathcal{T})$. The $i$th ($i=1, \ldots, N$) sample curve is observed at time points $\{t_{i,1}<t_{i,2}< \cdots< t_{i, T_i}\}$ in $\mathcal{T}$ with ${\rm E}\{\bm{\mathcal{Y}}(t)\}$ and ${\rm cov}\{\bm{\mathcal{Y}}(t), \bm{\mathcal{Y}}(s)\}$ finite for $t, s \in \mathcal{T}$. Assume that all Conditions (A)$-$(E) hold. Then, we have the following: 
 
 1) $\sup\limits_{\bm{X} \in \mathcal{C}^p(\mathcal{T})}|GMFID_N(\bm{X}; F_{\bm{\mathcal{Y}}, N}, \beta)-GMFID(\bm{X}; F_{\bm{\mathcal{Y}}}, \beta)| \xrightarrow[]{a.s.} 0$ as $T_s \to \infty$ and $N \to ~\infty$.
 
 2) $\sup\limits_{\bm{X} \in \mathcal{C}^p(\mathcal{T})}|GMFED_N(\bm{X}; F_{\bm{\mathcal{Y}}, N})-GMFED(\bm{X}; F_{\bm{\mathcal{Y}}})| \xrightarrow[]{a.s.} 0$ as $T_s \to \infty$ and $N \to \infty$.
 \label{consistency}
 \end{theorem}
 \vspace{-0.5cm}
With Theorem \ref{consistency}, we can use the finite sample depths to estimate their population definitions. When computing the sample LMFD$_N$s, the multivariate depths may not be accurate at some indexes because of the small number of observations; hence, they can be estimated with binwise multivariate depths. To compute the sample GMFD$_N$s, first, the binwise sample mean and sample covariance matrix are estimated. Subsequently, the multivariate depth of any normalized data with respect to a subset of normalized data across time indexes is calculated. The estimation procedure is described in Section \ref{estimate}.

\subsection{Estimation of Finite Sample Depths}
\label{estimate}
 We mainly focus on the estimation of multivariate depths before applying them to possible frameworks of MFDs. For this purpose, it is necessary to divide $\mathcal{T}^N$ into several bins such that the separation points satisfy that $t^N_j=G_N^{-1}(j/\overline{T})$ for $j = 0, \ldots, \overline{T}$. The bins have three usages. First, they are used to calculate the binwise depths in local MFDs. Second, they are used to calculate the binwise sample mean and covariance in global MFDs. Third, they can be used to specify the weights when they are proportional to the time density. The halfspace depth (\citeauthor{tukey1975mathematics} \citeyear{tukey1975mathematics}), as a multivariate depth satisfying Condition (A), is used as a multivariate depth in GMFD and LMFD frameworks.
 
 
Assume that we have $P_k=\{\bm{x}_1, \ldots, \bm{x}_{d_k}\}~(\sum_{k=1}^{\overline{T}} d_k = \sum_{i=1}^N T_i)$ observations in the $k$th bin $\mathcal{T}^{N}_{k}=[t^N_{k-1}, t^N_k]$ for $k=1, \ldots, \overline{T}$. The pointwise depths $D(\bm{X}(t); F_{\bm{\mathcal{Y}}(t)})$ in LMFD$_N$ can be approximated by calculating $D(\bm{X}(t); F_{\bm{\mathcal{Y}}(t)})$ with respect to $P_k$. In other words, we replace the pointwise depths with binwise depths. The complexity of the exact computation of halfspace depth is $O(d_k^{p-1}\log{d_k})$ according to \citeauthor{rousseeuw1998computing} (\citeyear{rousseeuw1998computing}) and \citeauthor{dyckerhoff2016exact} (\citeyear{dyckerhoff2016exact}), and it is implemented in $\overline{T}$ intervals.

Comparatively, to obtain the pointwise depths in GMFD$_N$, it is necessary to scale the data and combine the scaled data across the time index. To remove the influence of outliers, we perform robust principal component analysis (rPCA, \citeauthor{croux2007algorithms} \citeyear{croux2007algorithms}) to obtain the sample center and covariance. The aim of rPCA is to find the principal components $\bm{a}^{l_k}$~($1\leq l_k\leq p$) to maximize the variance of the projected data on $\bm{a}^{l_k}$. Then, let the scaled data in $\mathcal{T}^N_k$ be $P^*_k=\{\bm{x}^*_1, \ldots, \bm{x}^*_{d_k}\}$ for $k=1, \ldots, \overline{T}$, and let the combined scaled data $P^*=P^*_1 \cup \cdots\cup P^*_{\overline{T}}$. To accelerate the process of global pointwise halfspace depth computation (\citeauthor{cuevas2007robust} \citeyear{cuevas2007robust}), we consider the depth of a point $\bm{x}^* \in \mathbb{R}^p$ with respect to a subset of $P^*$. Let the cardinality of the subset $m=\sum_{i=1}^N{T_i}$ if $\sum_{i=1}^N{T_i} < 1000$ and $N_s = 1000$ otherwise. Then, we randomly select $N_s$ samples from $P^*$ without replacement and calculate the pointwise depth of $\bm{X}^*(t)$ $\forall t \in \mathcal{T}^N$ relative to the subset as an approximate estimation of the pointwise depth. The complexity becomes $O(N_s^{p-1}\log{N_s})$.

In summary, we have six depths; see Table \ref{depth_table}. There are three GMFDs: GMFID with the weight proportional to the time density at $t$ (GMFID $(w_t)$), GMFID with the weight proportional to the volume of the depth region at time $t$ (GMFID $(w_d)$), and GMFED. Further, there are three LMFDs: LMFID $(w_t)$, LMFID $(w_d)$, and LMFED. We will discuss the pros and cons of the above depths in the robustness of extracting the median, central region, and outliers under outlier contamination and time sparseness. In addition, the running time is also regarded as an assessment index. The performances of these six depths are explored in Section \ref{section3}.

\section{Simulation Study}
\label{section3}
\subsection{Model Setting}
Four bivariate models are built according to Eq. (\ref{decomposition_y}): $\bm{Y}(t)=\bm{\mu}(t) +\sum_{m=1}^{M} \rho_{m} \bm{\varphi}_{m}(t) + \bm{\epsilon}(t).$ We let $\mathcal{T}=[0, 1]$ and the time density function $g(t)$ be the uniform density on $t \in \mathcal{T}$. Then, we generate $N=200$ curves for $t \in \mathcal{T}$, $M \in \mathbb{Z}$. Additionally, let $\bm{\varphi}_m(t)$ be the $m$th Fourier basis function (\citeauthor{happ2018multivariate} \citeyear{happ2018multivariate}), $\rho_m \sim \mathcal{N}(0, e^{-\frac{m+1}{2}})$, and $\epsilon^{(l)}(t) \sim \mathcal{N}(0, \mathcal{U}(0, 0.1))$ for $l = 1, 2$, where $\mathcal{U}$ denotes the uniform distribution.

The four models have different $\bm{\mu}$, $\bm{\varphi}_m$, or $M$ values. \textbf{Model I}: $\bm{\mu}(t)=5 (\cos(2 \pi t), \sin(2 \pi t))^\top$, $M = 8$. \textbf{Model II}: $\bm{\mu}(t)=(-4t, 5t)^\top$, $M = 8$. \textbf{Model III}: $\bm{\mu}(t)=(4t, 6(t - 1/2)^2)^\top$, $M = 2$. \textbf{Model IV}: $\bm{\mu}(t)=(4t, 0)^\top$, $\rho_1 \sim \mathcal{U}(-7, 7)$, $\bm{\varphi}_1(t)=(0, t^2 - t)^\top$, $M = 1$.

In each model, a 10\% proportion of outliers can be introduced. Let $S = \{\bm{Y}_1, \ldots, \bm{Y}_N\}$ be the set of samples, and $O$ be the set of introduced outliers out of $S$. When there are no outliers, we regard the samples $\{\bm{Y}\}^N_{i=1}$ to be from the stochastic process with cdf $F_{\bm{\mathcal{Y}}}$. In contrast, when there are outliers, we regard the contaminated samples $\{\bm{Y}_c\}^N_{i=1}$ to be from the stochastic process with cdf $F_{\bm{\mathcal{Y}}_c}$. The outlier types we consider are magnitude outliers I and II, amplitude outliers I and II, and shape outliers I and II; see below. 

For any above model, we obtain the maximum $\overline{\bm{m}}: = (\overline{m}^{(1)}, \overline{m}^{(2)})^\top$ and the range $\bm{r}:=(r^{(1)}, r^{(2)})^\top$ of data generated from the standard model. Let $s$ take the value 1 with probability $0.5$ and the value $-1$ with probability $0.5$. The magnitude outliers are shift outliers in the whole or in a part of the time domain. Then, a magnitude outlier I is $\bm{Y}_c(t)=\bm{Y}(t) +  as\bm{r}$, with $a \sim \mathcal{U}(0.8, 1)$. Similarly, a magnitude outlier II is $\bm{Y}_c(t)=\bm{Y}(t) + as\bm{r}$, where $t = t' + 0.1$, and $t' \sim \mathcal{U}(0, 0.85)$. The amplitude outliers are patterns with similar shapes but different amplitudes. For example, an amplitude outlier I is $\bm{Y}_c(t)= (1 + a)s\bm{Y}(t)$ and an amplitude outlier II is $\bm{Y}_c(t)= (1 - a)s\bm{Y}(t)$. 

\begin{figure}[!ht]
\centering
\includegraphics[width = 0.32\textwidth, height = 4.5cm]{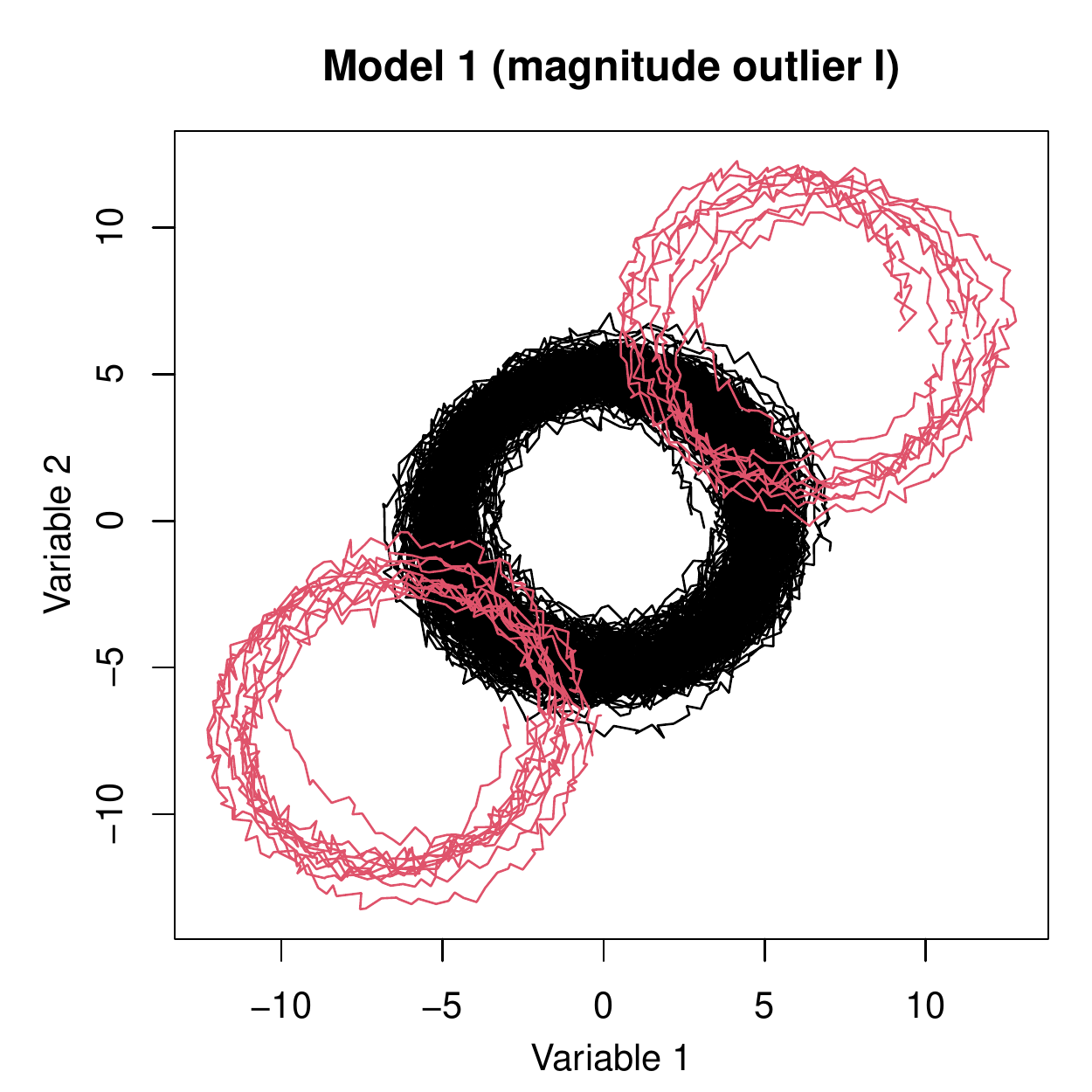}
\includegraphics[width = 0.32\textwidth, height = 4.5cm]{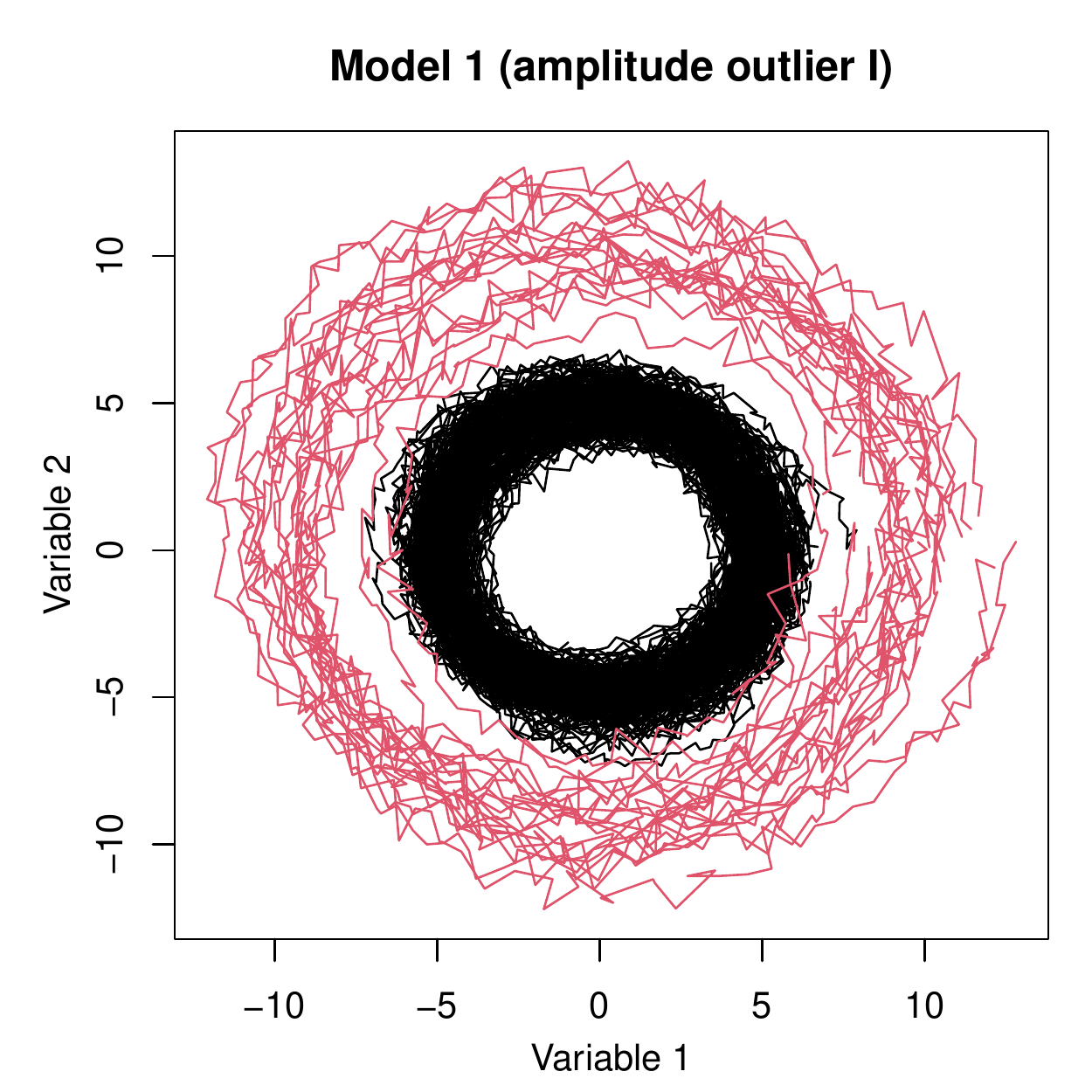}
\includegraphics[width = 0.32\textwidth, height = 4.5cm]{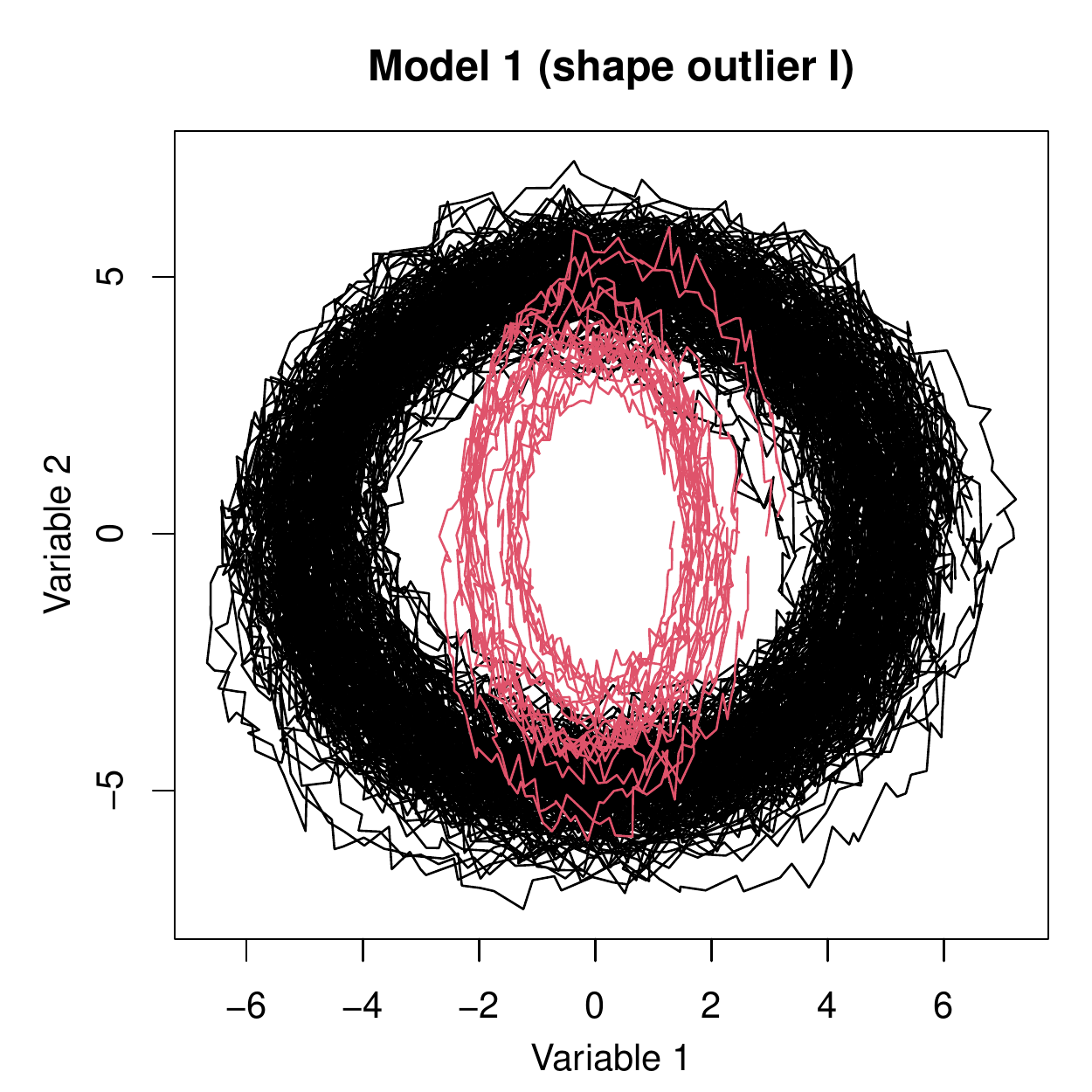}
\includegraphics[width = 0.32\textwidth, height = 4.5cm]{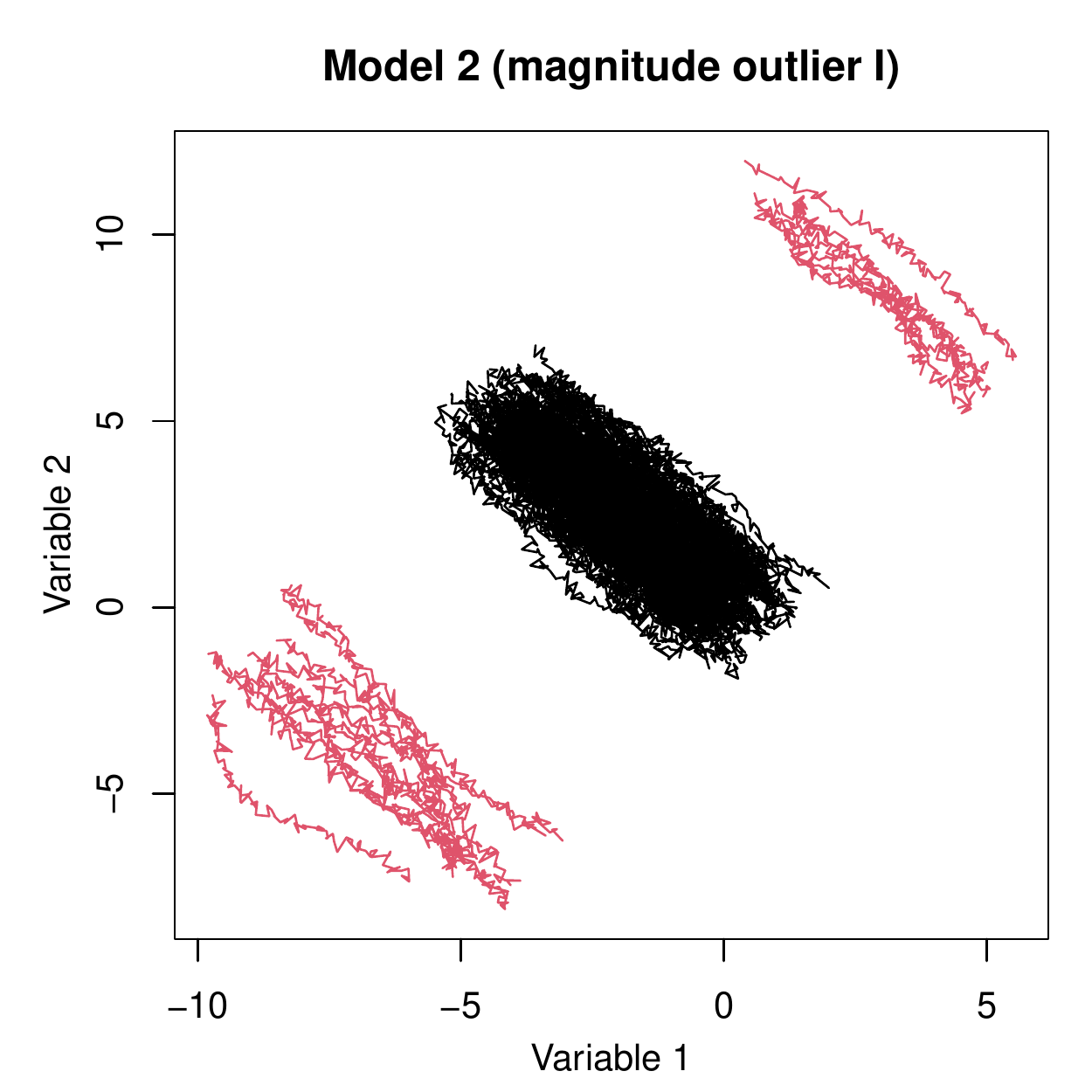}
\includegraphics[width = 0.32\textwidth, height = 4.5cm]{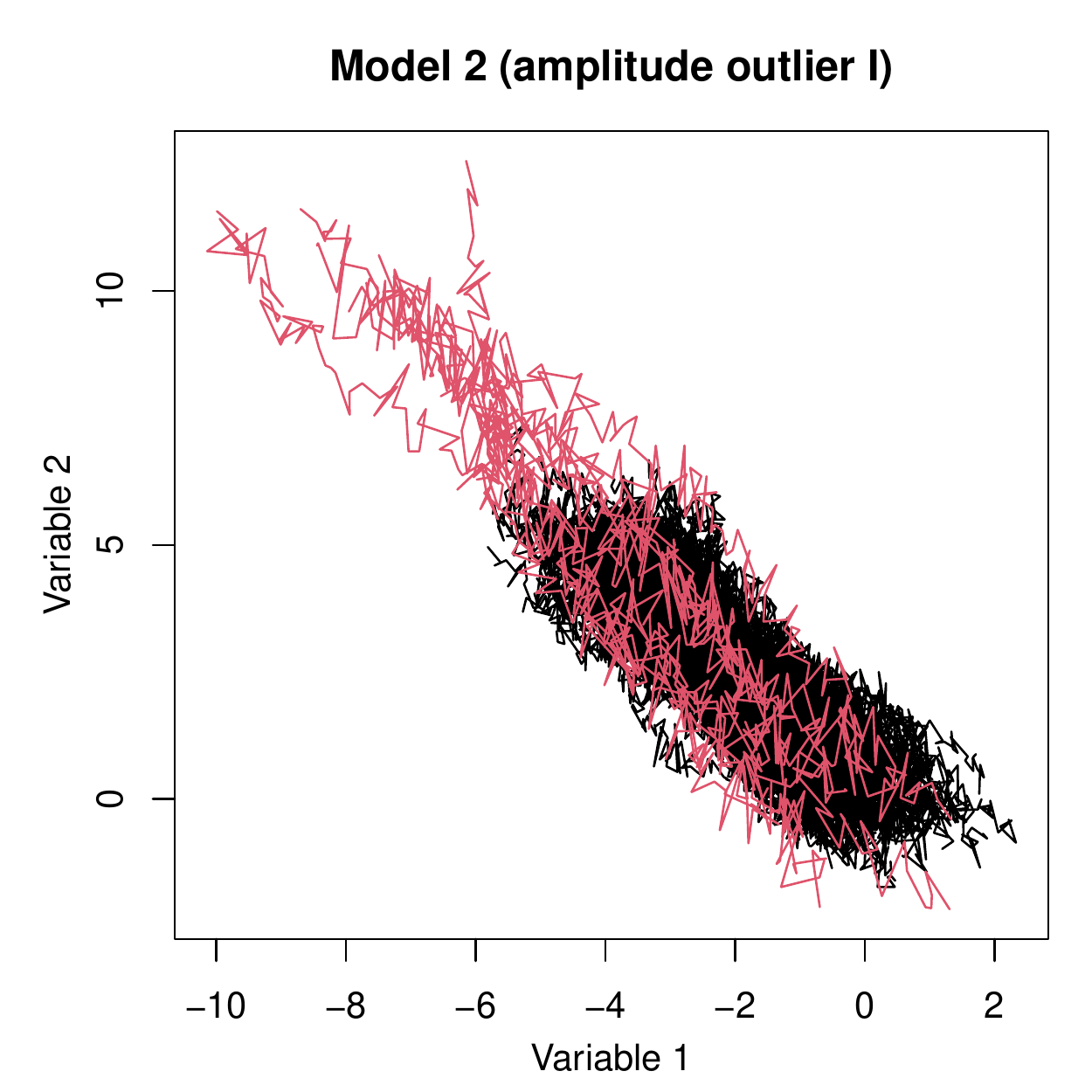}
\includegraphics[width = 0.32\textwidth, height = 4.5cm]{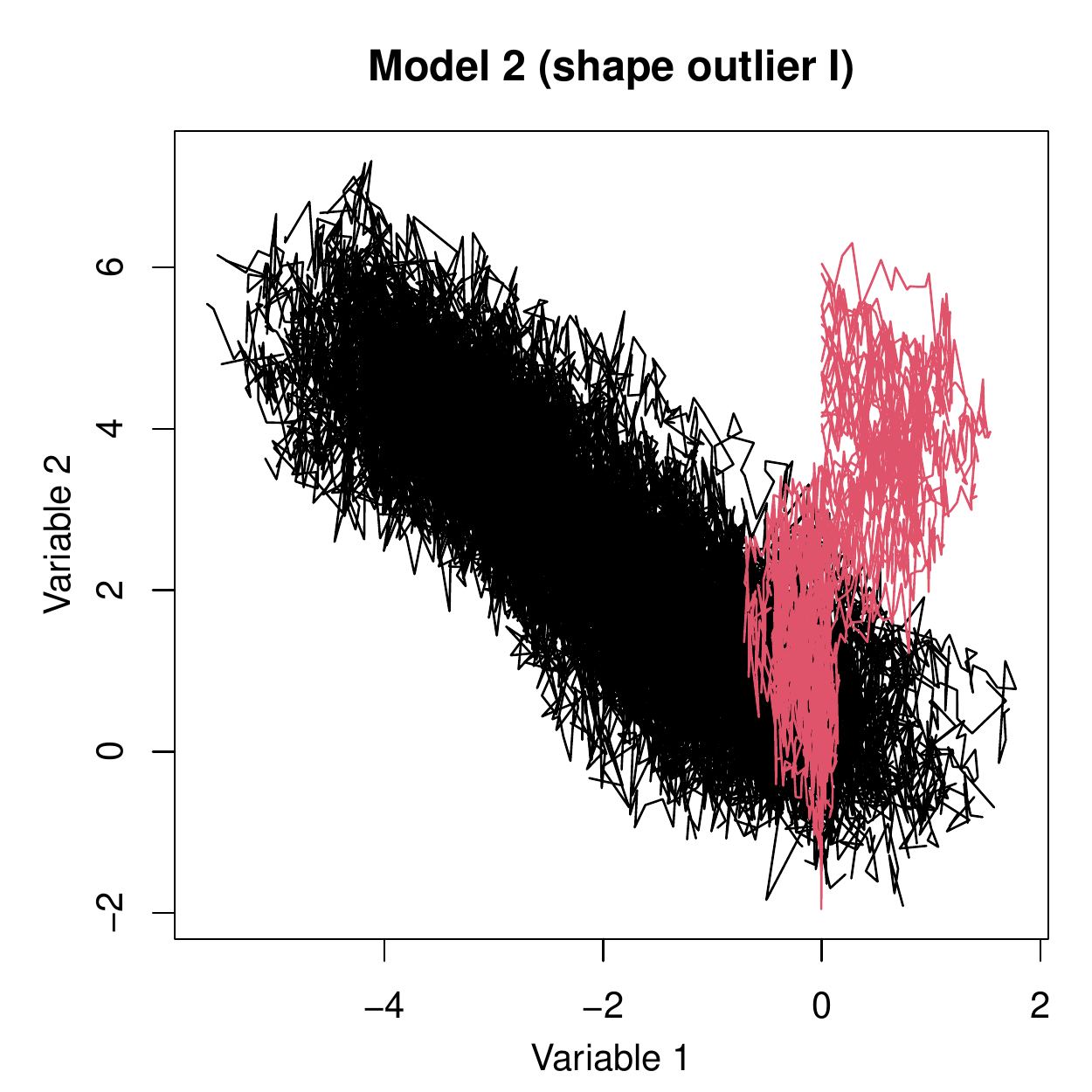}
\includegraphics[width = 0.32\textwidth, height = 4.5cm]{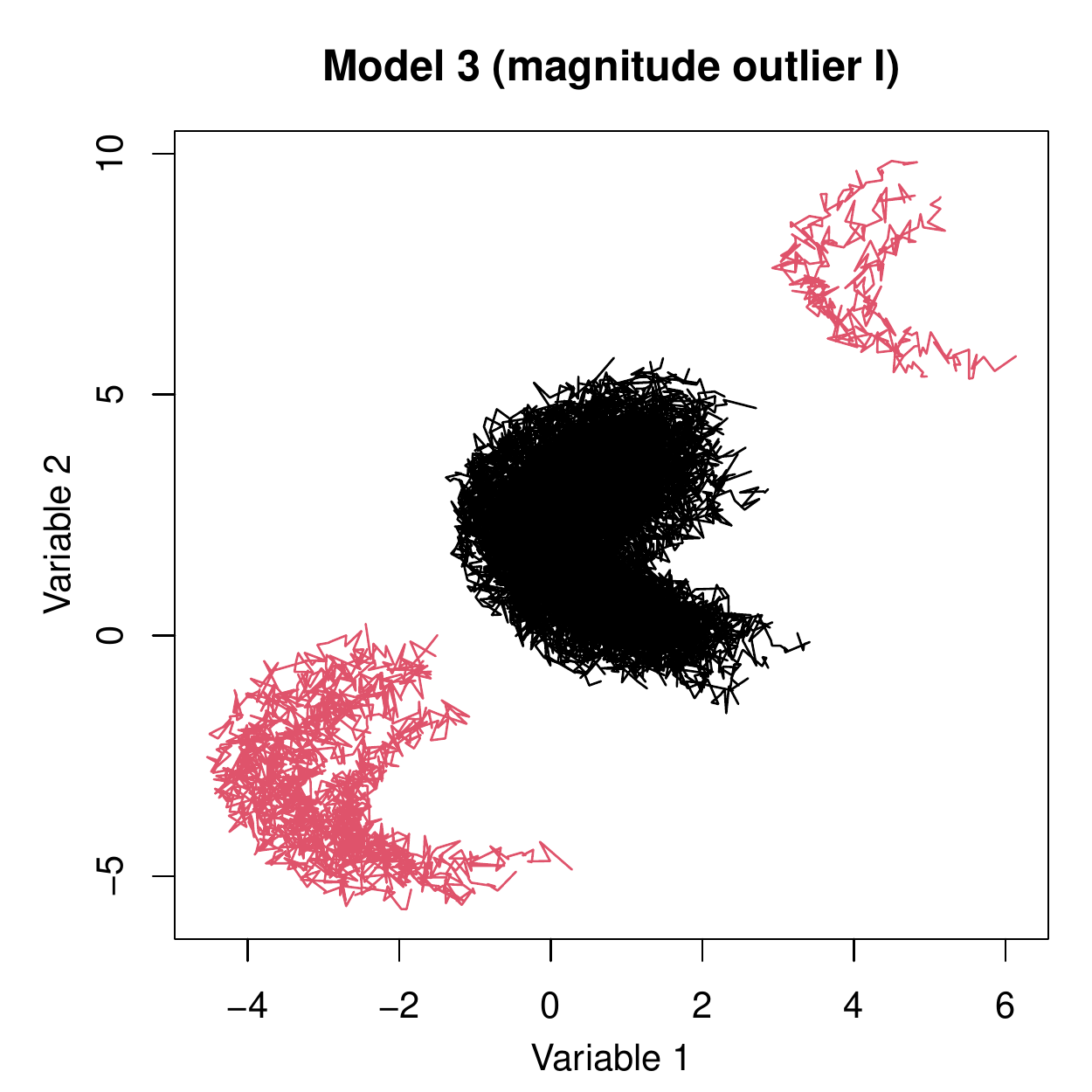}
\includegraphics[width = 0.32\textwidth, height = 4.5cm]{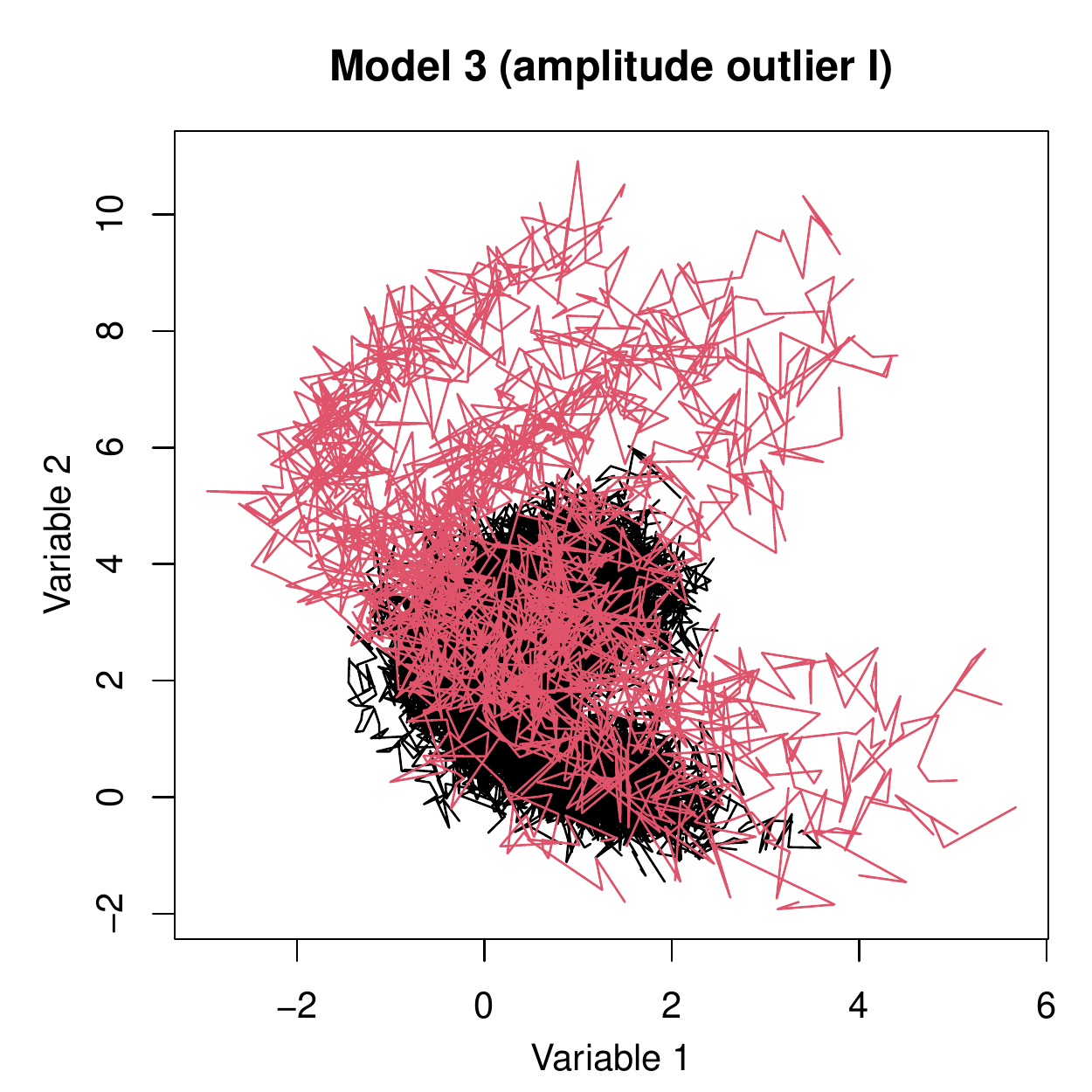}
\includegraphics[width = 0.32\textwidth, height = 4.5cm]{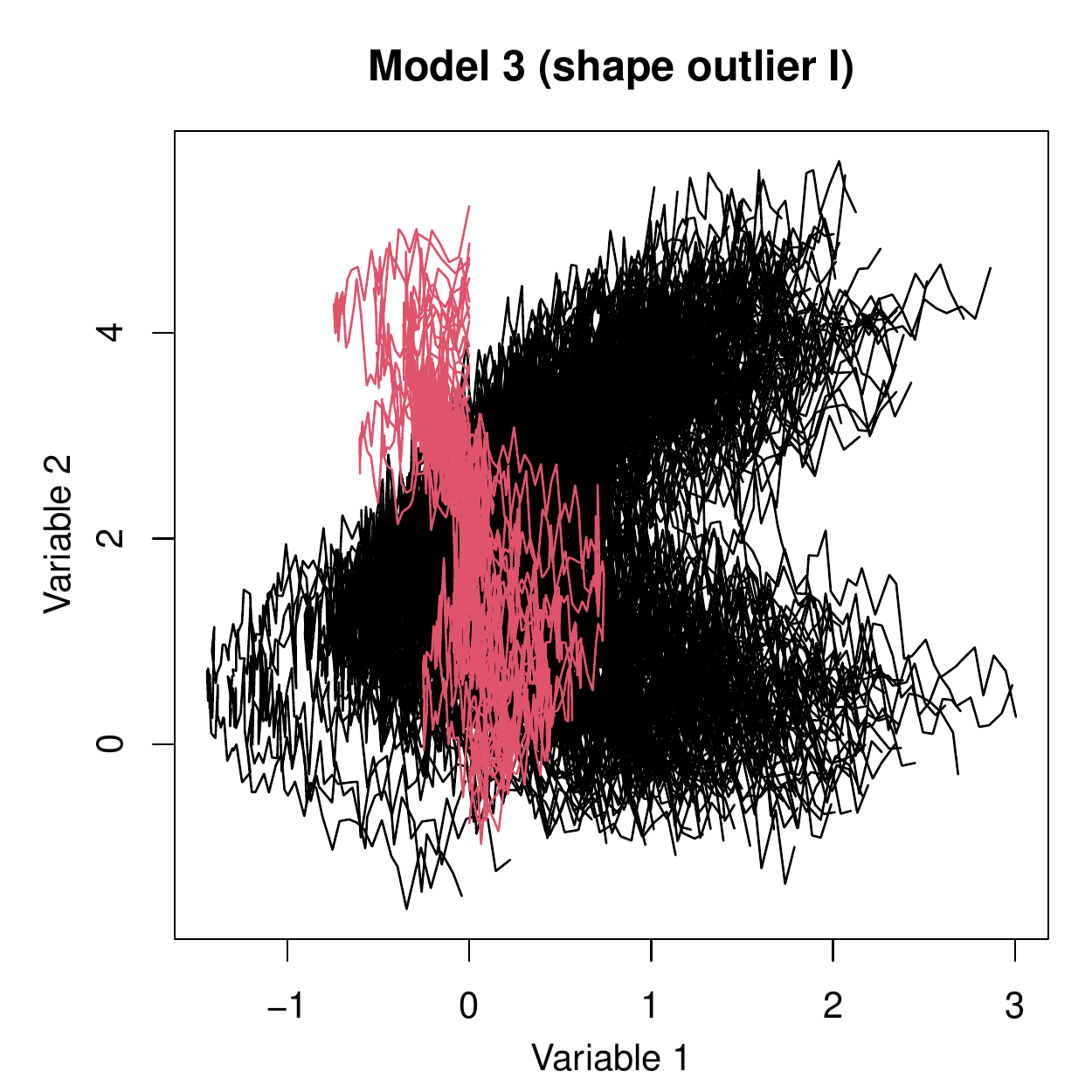}
\includegraphics[width = 0.32\textwidth, height = 4.5cm]{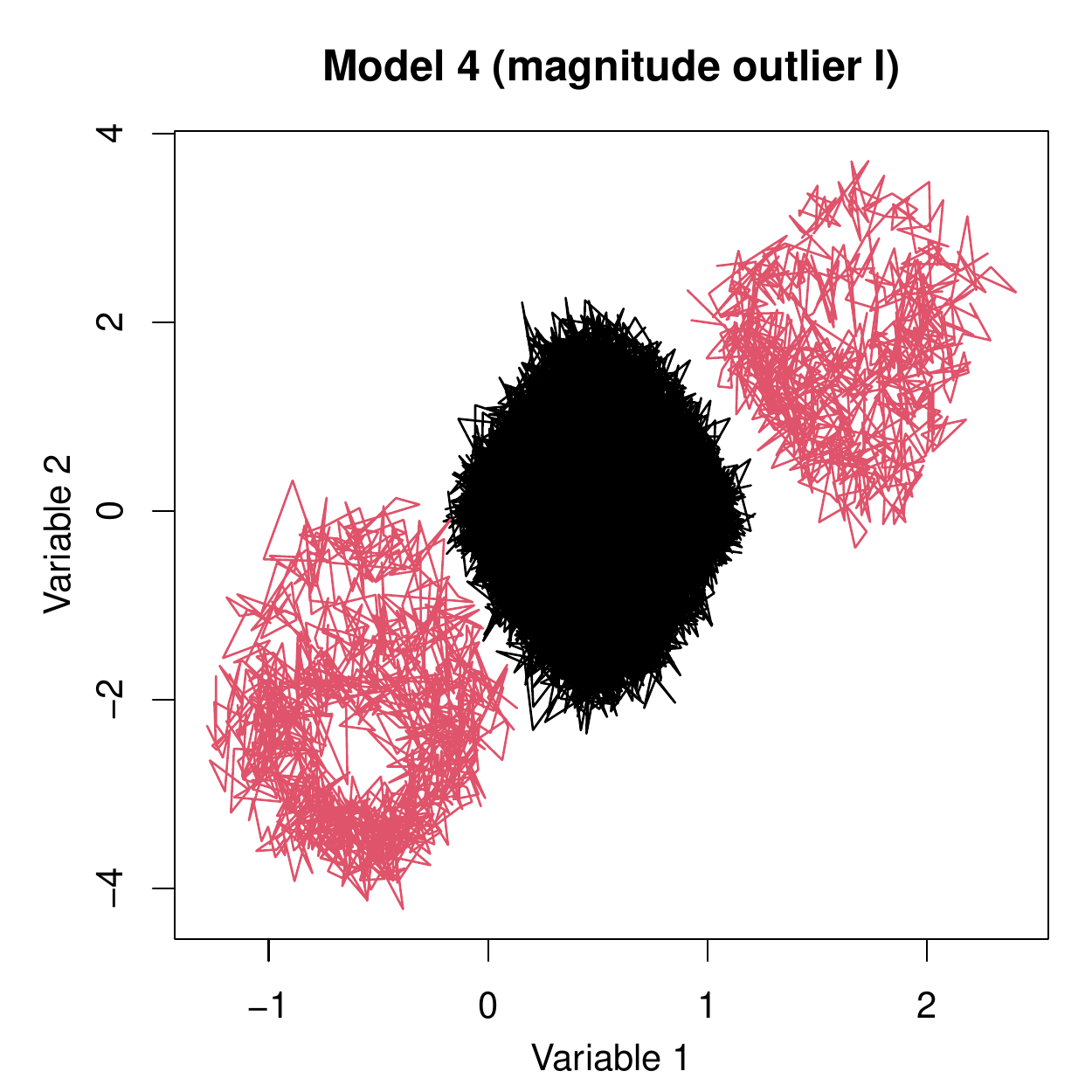}
\includegraphics[width = 0.32\textwidth, height = 4.5cm]{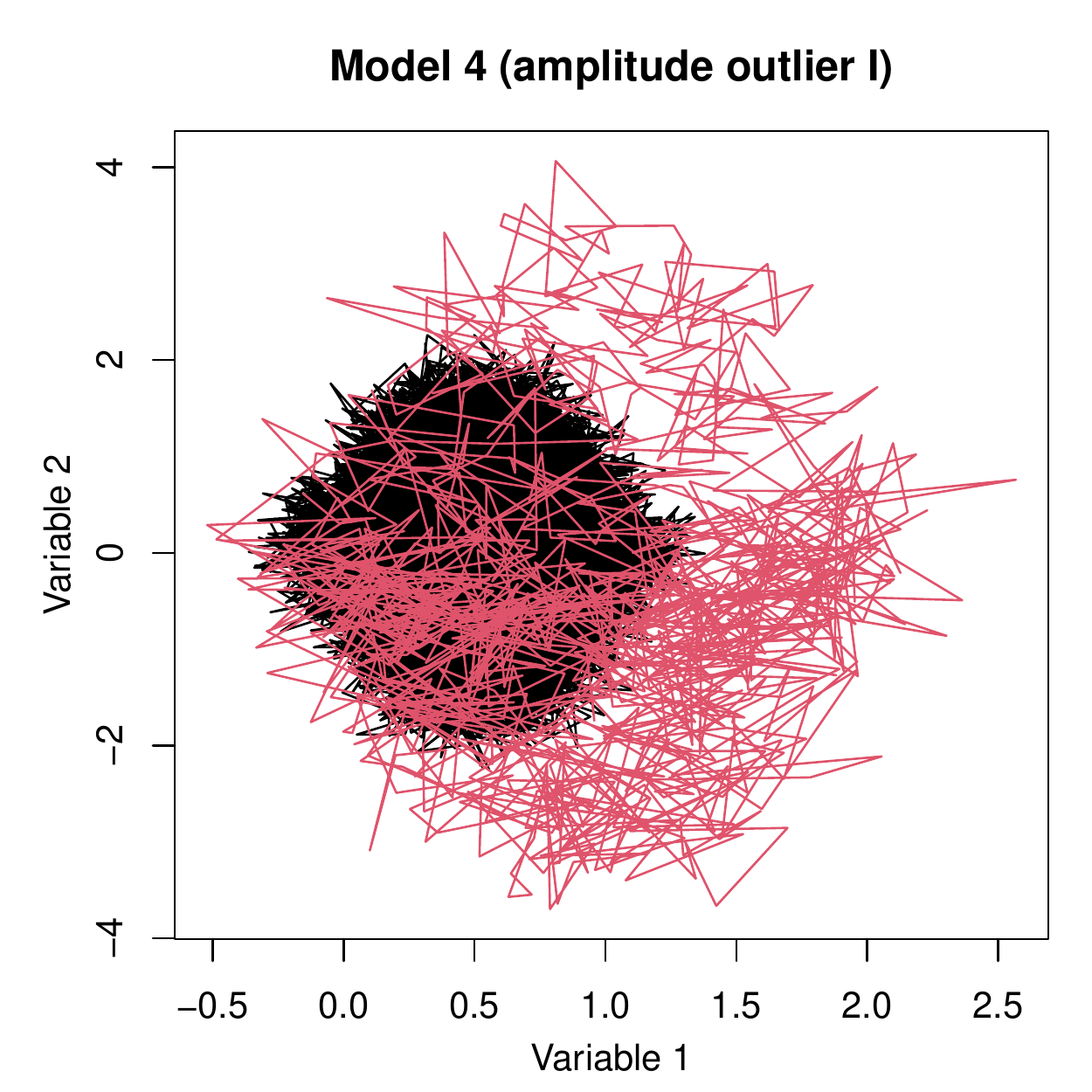}
\includegraphics[width = 0.32\textwidth, height = 4.5cm]{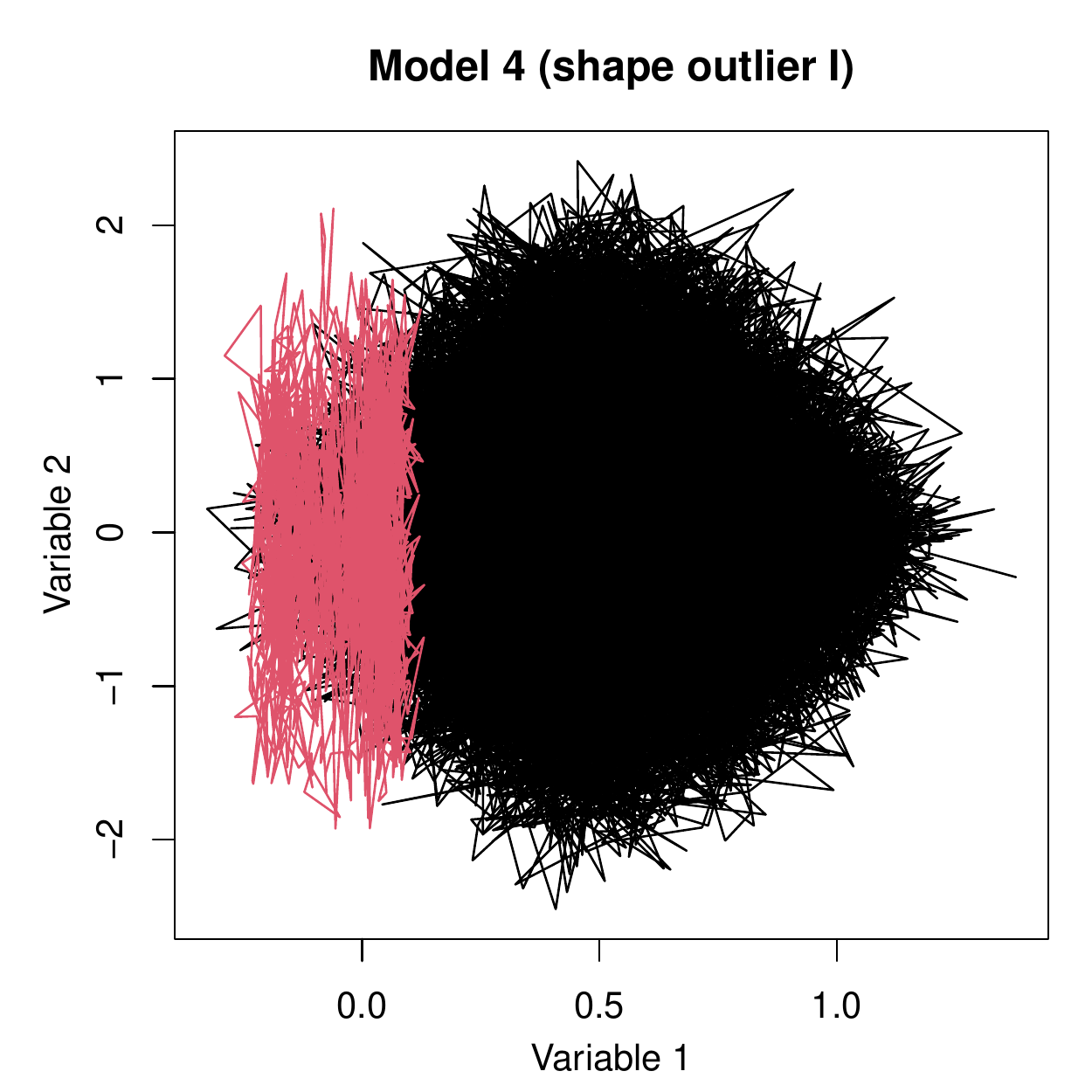}
\caption{Panels from top to bottom show Models 1$-$4 contaminated by different types of outliers. Each panel shows a model contaminated by a 10\% proportion of magnitude outlier I, amplitude outlier I, and shape outlier I, respectively, from left to right. The non-outlying and outlying curves are colored in black and red, respectively.}
\label{simulation_visualization}
\end{figure}

Naturally, the shape outliers are curves that do not resemble the bulk of curves in the standard model.~For Model 1, a shape outlier I is $\bm{Y}_c(t)= (a_1{Y}^{(1)}(t), 2a_1{Y}^{(2)}(t))^\top$ for $a_1 \sim \mathcal{U}(0.3, 0.5)$; and a shape outlier II is $Y^{(1)}_c(t) = a_2 Y^{(1)}_c(t)$ for $a_2 \sim \mathcal{U}(1.6, 1.8)$.
For Models 2$-$4, a shape outlier I is ${Y}^{(j)}_c(t)= Y^{(j)}(t) + \frac{1}{5}a\overline{m}^{(j)}\sin(2\pi t)$ for $j = 1, 2$; and a shape outlier II is $\bm{Y}_c(t)= a\bm{Y}_c(t) + a(\cos(\pi t), \sin(\pi t))^\top$ with $a \sim \mathcal{U}(0.8, 1)$.
Models 1$-$4 with magnitude outlier I, amplitude outlier I, and shape outlier I are shown in Figure \ref{simulation_visualization}.

In addition, sparseness types, namely, point, peak, and partial sparseness, are considered. We use $p_s$ and $p_{curve}$ as the sparseness-tuning parameters, where $p_s$ represents the proportion of curves with at least one missing value among the whole samples, and $p_{curve}$ represents the proportion of missing indexes throughout the grid. With regard to the extent of sparseness, we let $p_{curve} = 0\%$ represent the case without sparseness; $p_{curve}\sim\mathcal{U}(0.1, 0.3)$ represents the case with medium sparsity, and $p_{curve}\sim\mathcal{U}(0.4, 0.6)$ that with high sparsity. 

For any curve $\bm{Y}_{c,i}~(i=1, \ldots, N)$, we assume that its observed time points ${t}_{o,i}$ are defined by
\[{t}_{o,i}=
\begin{cases}
\mathcal{U}(0, 1), & \textrm{point~sparseness}, \\
x\mathcal{U}(0, t_{i, s}) + (1 - x)\mathcal{U}(t_{i, s} + p_{curve}, 1) , & \textrm{peak~sparseness}, \\
\mathcal{U}(0, t_{i, s}), & \textrm{partial~sparseness},
\end{cases}\]
where $x\sim \mathcal{B}(0.5)$, and $\mathcal{B}$ represents the Bernoulli distribution. We let $t_{i,s} \sim \mathcal{U}(0, 0.7 - p_{curve})$ (and $t_{i,s} \sim \mathcal{U}(0, 1 - p_{curve})$) if the data show peak sparseness (and partial sparseness).

There are two differences between the sparseness settings in this study and those in the work of  \citeauthor{qu2022sparse} (\citeyear{qu2022sparse}). First, the peak sparseness type here allows various missing initial locations $t_{i,s}$ per subject. Second, we assume that data are either observed or missing simultaneously across components of each curve because we propose multivariate functional depths based on multivariate depths. 

\subsection{Assessment Criteria}
The candidate MFDs include the six depths we summarized: GMFID $(w_t)$, GMFID $(w_d)$, GMFED, LMFID $(w_t)$, LMFID $(w_d)$, and LMFED. We take $\beta = 1/4$ when $w$ is proportional to the volume of the depth region. Besides, we add two comparative depths: the multivariate partially observed integrated functional depth (MPOIFD, \citeauthor{elias2022integrated} \citeyear{elias2022integrated}) with time-varying weights, and the multivariate functional halfspace depth applied to reconstructed data from the multivariate functional principal component analysis (MFHD$_{mfpca}$, see \citeauthor{claeskens2014multivariate} \citeyear{claeskens2014multivariate}, \citeauthor{qu2022sparse} \citeyear{qu2022sparse}). 
A summary of the aforementioned eight depths is shown in Table \ref{depth_table} with their references and types of time grids.
\begin{table}[!ht]
\caption{Summary of the candidate multivariate functional depths in the simulation study.}
\label{depth_table}
\begin{scriptsize}
\begin{center}
\vspace{-0.5cm}
\begin{tabular}{||c| c| c| c| c |c||} 
 \hline
 \begin{tabular}{@{}c@{}}Depth \\ Abbreviation\end{tabular} & Depth name & Reference & Weight & \begin{tabular}{@{}c@{}@{}} To Irregularly \\ Observed Data \\ on Various Grids \end{tabular} & \begin{tabular}{@{}c@{}@{}} To Partially \\ Observed Data\\ on Common Grids \end{tabular} \\ [0.5ex] 
 \hline\hline
 GMFID ($w_t$) & \begin{tabular}{@{}c@{}}global multivariate \\ functional integrated depth\end{tabular}  & this work & time & \textcolor{green}{\CheckmarkBold} & \textcolor{green}{\CheckmarkBold}\\ 
 \hline
 GMFID ($w_d$)& \begin{tabular}{@{}c@{}}global multivariate \\ functional integrated depth\end{tabular} & this work & depth region & \textcolor{green}{\CheckmarkBold} & \textcolor{green}{\CheckmarkBold}\\
 \hline
 GMFED & \begin{tabular}{@{}c@{}}global multivariate \\ functional extremal depth\end{tabular} & this work & none & \textcolor{green}{\CheckmarkBold} & \textcolor{green}{\CheckmarkBold}\\
 \hline
 LMFID ($w_t$) & \begin{tabular}{@{}c@{}}local multivariate \\ functional integrated depth\end{tabular} & this work&time  &\textcolor{green}{\CheckmarkBold} & \textcolor{green}{\CheckmarkBold}\\
 \hline
 LMFID ($w_d$) & \begin{tabular}{@{}c@{}}local multivariate \\ functional integrated depth\end{tabular} & this work&depth region &  \textcolor{green}{\CheckmarkBold}& \textcolor{green}{\CheckmarkBold} \\ [1ex] 
 \hline
 LMFED & \begin{tabular}{@{}c@{}}local multivariate \\ functional extremal depth\end{tabular} & this work & none & \textcolor{green}{\CheckmarkBold} & \textcolor{green}{\CheckmarkBold}\\
 \hline
 MPOIFD & \begin{tabular}{@{}c@{}@{}}multivariate partially\\ observed integrated \\ functional depth\end{tabular} &
 \begin{tabular}{@{}c@{}} \citeauthor{elias2022integrated}  \\ (\citeyear{elias2022integrated}) \end{tabular} & time & \textcolor{red}{\XSolidBold} & \textcolor{green}{\CheckmarkBold}\\ [1ex] 
 \hline
 MFHD$_{mfpca}$ & \begin{tabular}{@{}c@{}}multivariate functional\\ halfspace depth\end{tabular} & \begin{tabular}{@{}c@{}} \citeauthor{qu2022sparse} \\ (\citeyear{qu2022sparse}) \end{tabular}  & depth region & \textcolor{red}{\XSolidBold} & \textcolor{red}{\XSolidBold}\\
 \hline
\end{tabular}
\end{center}
 \end{scriptsize}
 \vspace{-0.5cm}
 \end{table}
 
In addition, we explored another recommended depth, MFHD$_{bmfpca}$ (\citeauthor{qu2022sparse} \citeyear{qu2022sparse}), which is applied to reconstructed data from the bootstrap MFPCA. However, we decided not to include it in this comparison as it showed slight improvement in the rank-based Spearman correlation (\citeauthor{kendall1945treatment} \citeyear{kendall1945treatment}) with a relatively long running time. 

For assessing the robustness of the MFDs under outlier contamination and time sparseness, the median, central region, outliers, the ranking association for the nonoutliers, and the running time should be extracted.

1. \textbf{ASE median}: It is the averaged squared scaled error between the true and estimated central curve $\frac{1}{pT_j}\sum_{l=1}^p\sum_{j=1}^{T_j}\bigg(\frac{m^{(l)}_{\bm{\mathcal{Y}}_c}(\widetilde{t}_j) - m^{(l)}_{\bm{\mathcal{Y}}}(\widetilde{t}_j)}{R^{(l)}_{0.5}(\widetilde{t}_j)}\bigg)^2,$ where $\bm{m}_{\bm{\mathcal{Y}}}=(m^{(1)}_{\bm{\mathcal{Y}}}, \ldots, m^{(p)}_{\bm{\mathcal{Y}}})^\top$ and $\bm{m}_{\bm{\mathcal{Y}}_c}$ are the central curves from the noncontaminated model and contaminated model, respectively. Here, $\bm{m}_{\bm{\mathcal{Y}}}$ is estimated from the cross-sectional average of all nonoutliers, and $\bm{m}_{\bm{\mathcal{Y}}_c}$ is obtained based on the corresponding MFD with $\widetilde{t}_1, \ldots, \widetilde{t}_{T_j}$ time points. Let $R^{l}_{0.5}(\widetilde{t}_j)$ be the interquartile range of $\bm{\mathcal{Y}}$. It can be obtained from the cross-sectional range of the 50\% deepest curves by the corresponding MFD in the standard model.

2. \textbf{ASE central region}: It is the average squared error between the logarithm of the 50\%-dispersion curves computed on the contaminated and the uncontaminated data $\frac{1}{pT_j}\sum_{l=1}^p\sum_{j=1}^{T_j}\bigg\{\log\bigg(\frac{R^{(l)}_{c,0.5}(\widetilde{t}_j)}{R^{(l)}_{0.5}(\widetilde{t}_j)}\bigg)\bigg\}^2,$ where $R^{(l)}_{c,0.5}(\widetilde{t}_j)$ is the interquartile range of the $l$th variable from the $N$ samples in the contaminated model. It can be obtained from the 50\% deepest curves by the corresponding MFD in the contaminated model.

3. \textbf{The proportion of outliers lying in the 10\% least depth region}: $\frac{1}{\lfloor N/10 \rfloor}\sum\limits_{\bm{X} \in O}\bm{1}\{\bm{X}\\ \in \bm{Y}_{[r]}, \lceil 9N/10 \rceil < r \leq N\}$, where $\bm{Y}_{[r]}$ represents the $r$th deepest curve in $S$, $\lfloor N/10 \rfloor$ represents the biggest number not greater than $N/10$, and $\lceil 9N/10 \rceil$ represents the the smallest number not smaller than $9N/10$.

4. \textbf{The Spearman correlation of the rank-based variables} $\bm{r}_{\bm{\mathcal{Y}}}$ and $\bm{r}_{\bm{\mathcal{Y}}_c}$: $\rho_s = \frac{(\bm{r}_{\bm{\mathcal{Y}}}- \overline{r}_{\bm{\mathcal{Y}}}\bm{1}_{\lceil 9N/10 \rceil})^\top (\bm{r}_{\bm{\mathcal{Y}}_c}- \overline{r}_{\bm{\mathcal{Y}}_c}\bm{1}_{\lceil 9N/10 \rceil})}{\sqrt{\|\bm{r}_{\bm{\mathcal{Y}}}- \overline{r}_{\bm{\mathcal{Y}}}\bm{1}_{\lceil 9N/10 \rceil}\|^2 \|\bm{r}_{\bm{\mathcal{Y}}_c}- \overline{r}_{\bm{\mathcal{Y}}_c} \bm{1}_{\lceil 9N/10 \rceil}\|^2} }$,
where $\bm{r}_{\bm{\mathcal{Y}}}$ represents the rank of the nonoutliers in $\{\bm{Y}_{c,i}\}^N_{i=1}$, and $\overline{r}_{\bm{\mathcal{Y}}}$ represents the average rank of the corresponding group of data. Similarly, $\bm{r}_{\bm{\mathcal{Y}}}$ represents the rank of the aforementioned nonoutliers in $\{\bm{Y}_{c,i}\}^N_{i=1}$, $\overline{r}_{\bm{\mathcal{Y}}_c}$ represents the average rank of the corresponding nonoutliers in $\{\bm{Y}_{c,i}\}^N_{i=1}$, and $\bm{1}_{\lceil 9N/10 \rceil}$ represents a unit vector with length $\lceil 9N/10 \rceil$.

5. \textbf{The running time of the depth calculation}: It is the running time for each depth applied to the model with contamination and time sparseness. 

Let the simulation experiment run 100 times. On average, we expect to see low ASE measures close to 0, an outlier proportion lying in the small depth region and Spearman correlation close to 100\%, and the shortest running time possible. We compute the logarithm of the criterion for the central region. 

\subsection{Simulation Results}
\label{simulation_result}
\begin{figure}[!b]
\centering
\includegraphics[width = 0.49\textwidth, height = 4cm]{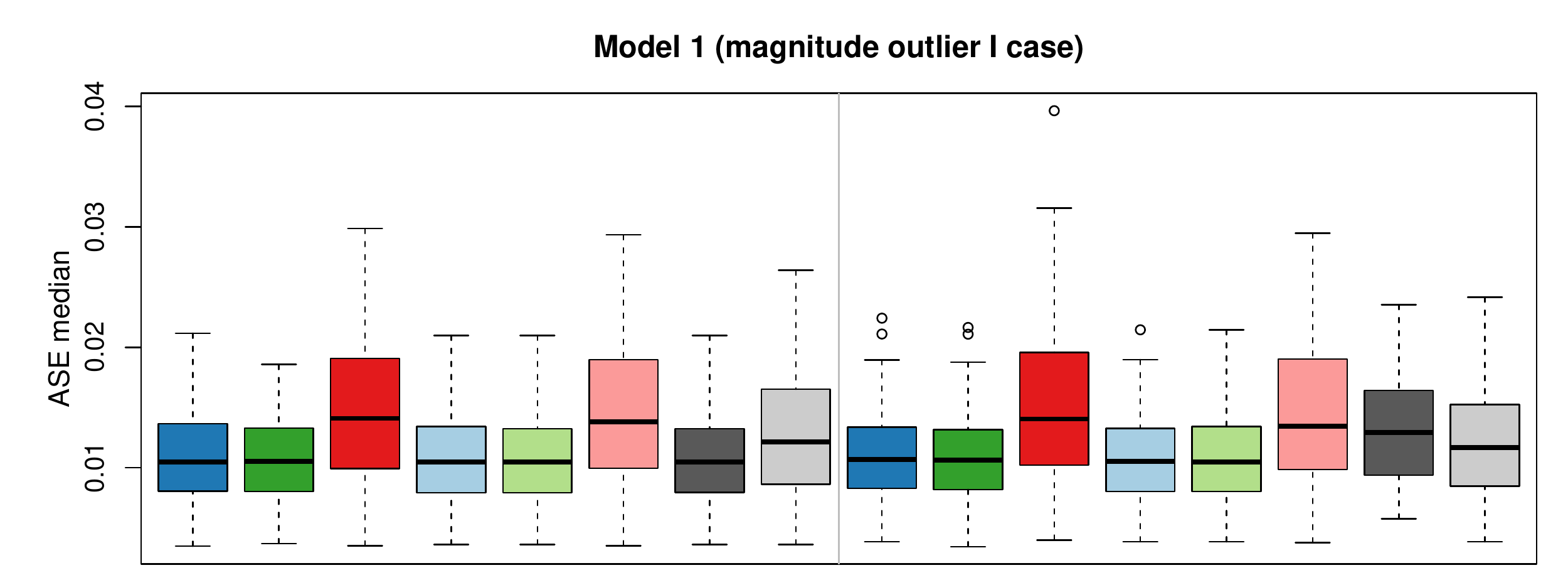}
\includegraphics[width = 0.49\textwidth, height = 4cm]{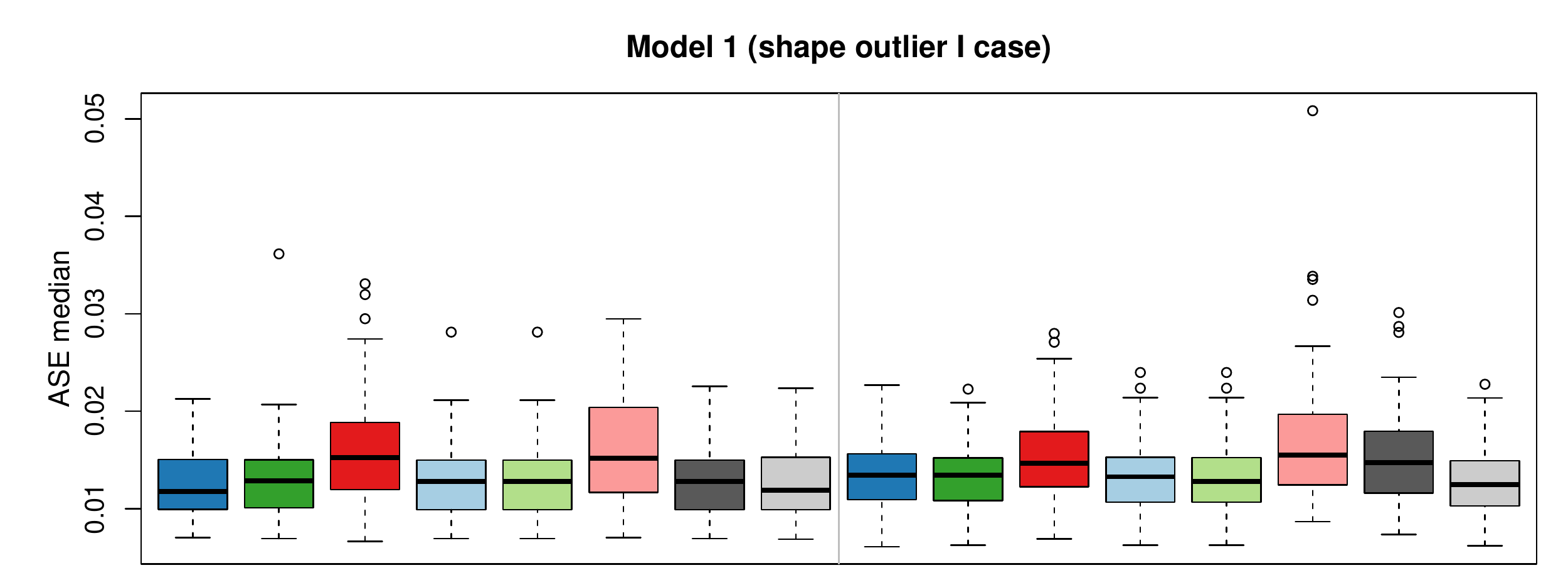}
\includegraphics[width = 0.49\textwidth, height = 3.5cm]{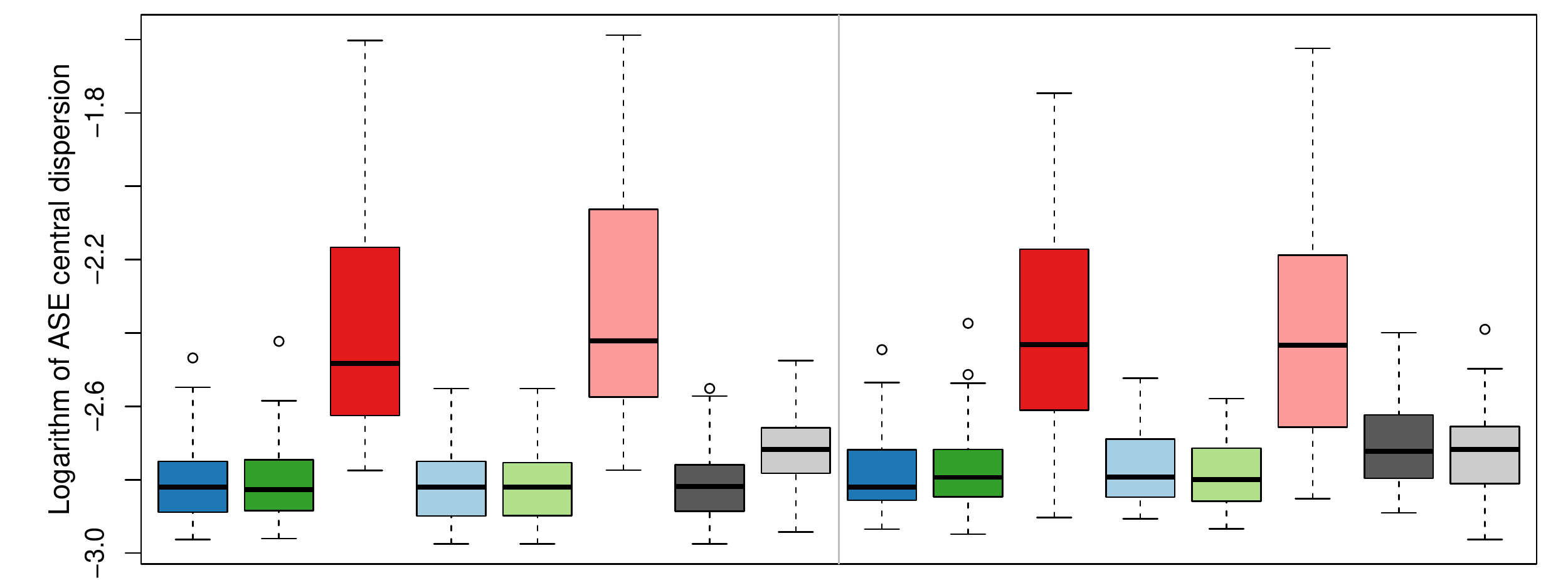}
\includegraphics[width = 0.49\textwidth, height = 3.5cm]{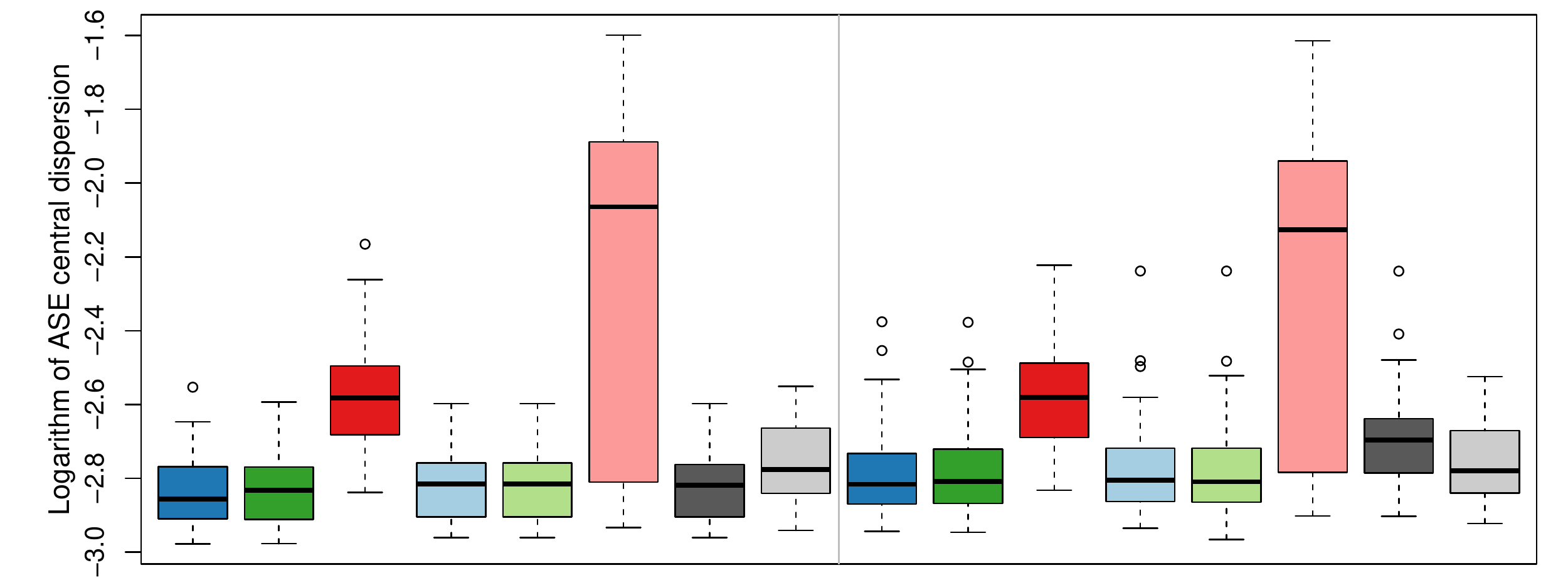}
\includegraphics[width = 0.49\textwidth, height = 3.5cm]{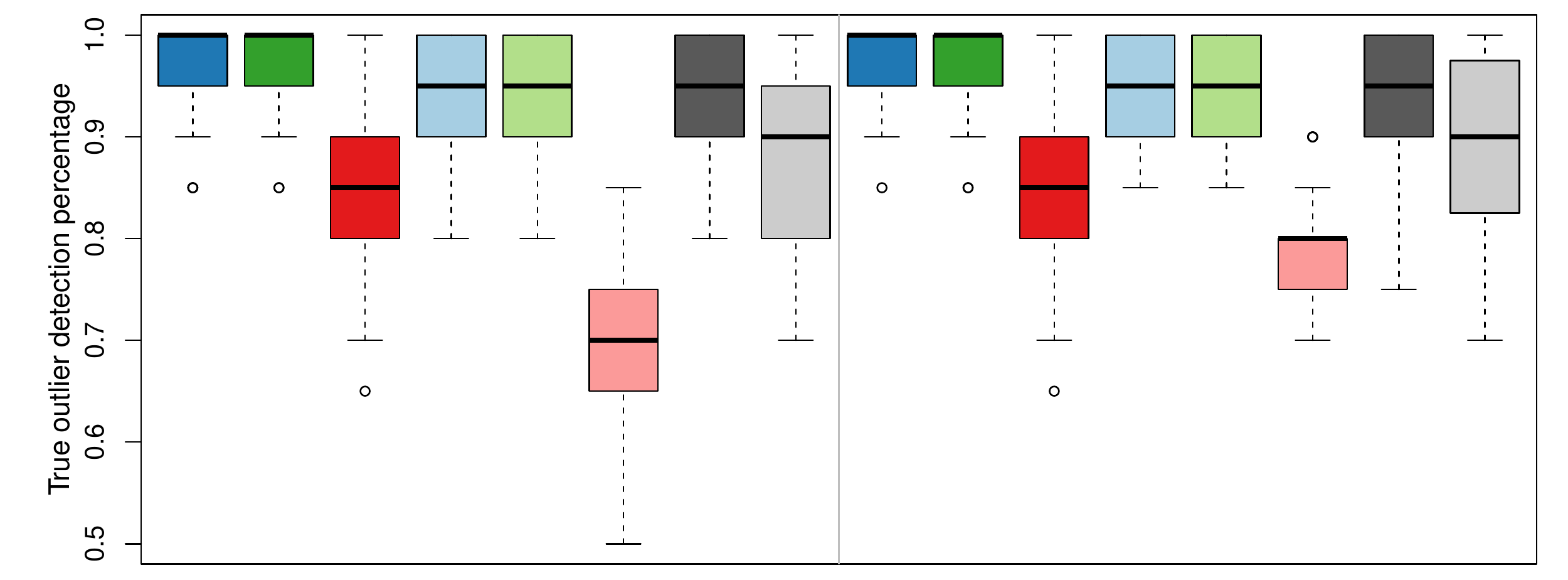}
\includegraphics[width = 0.49\textwidth, height = 3.5cm]{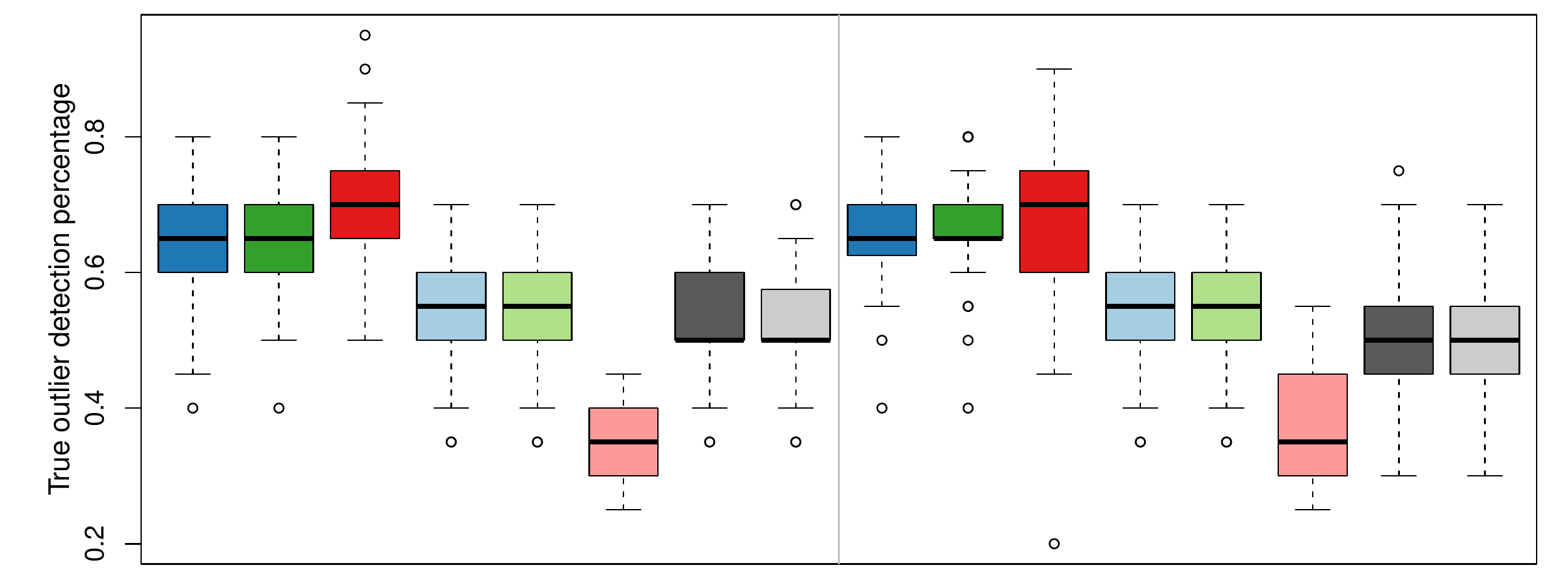}
\includegraphics[width = 0.49\textwidth, height = 4.2cm]{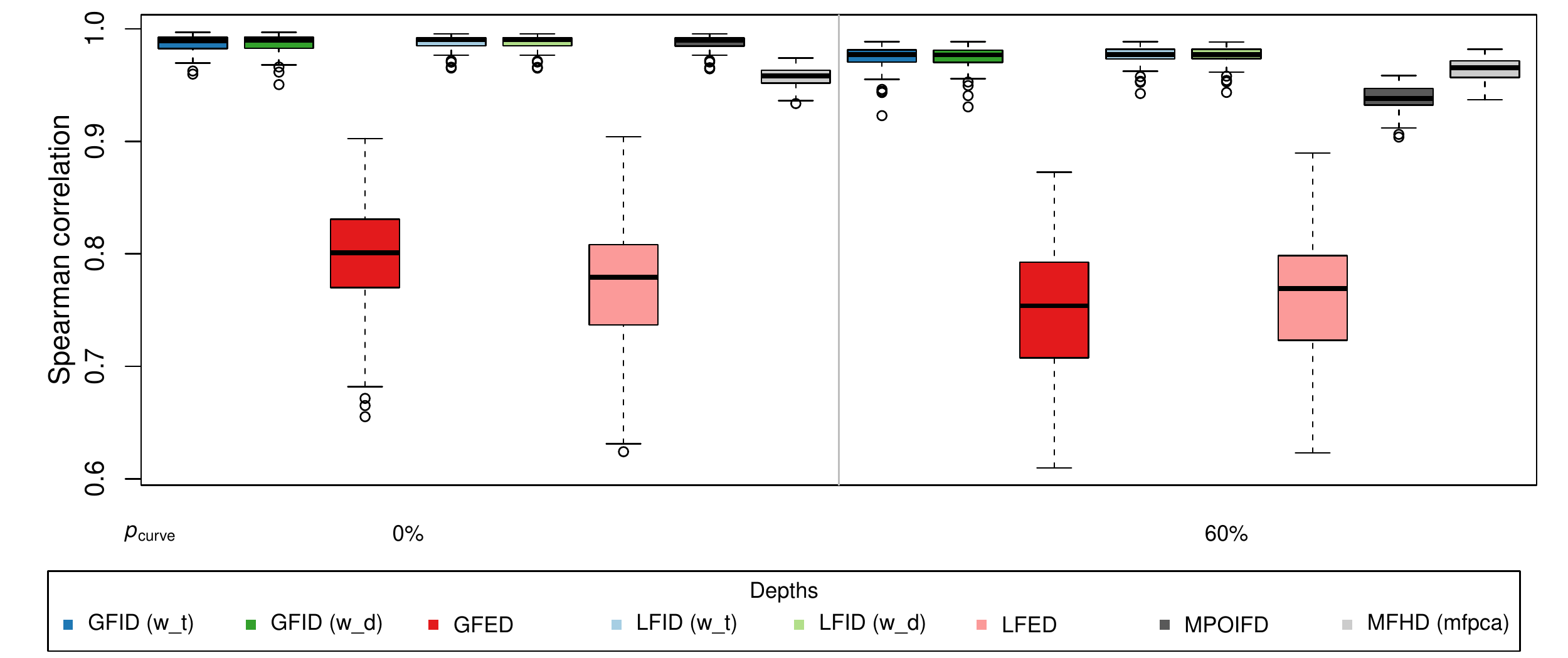}
\includegraphics[width = 0.49\textwidth, height = 4.2cm]{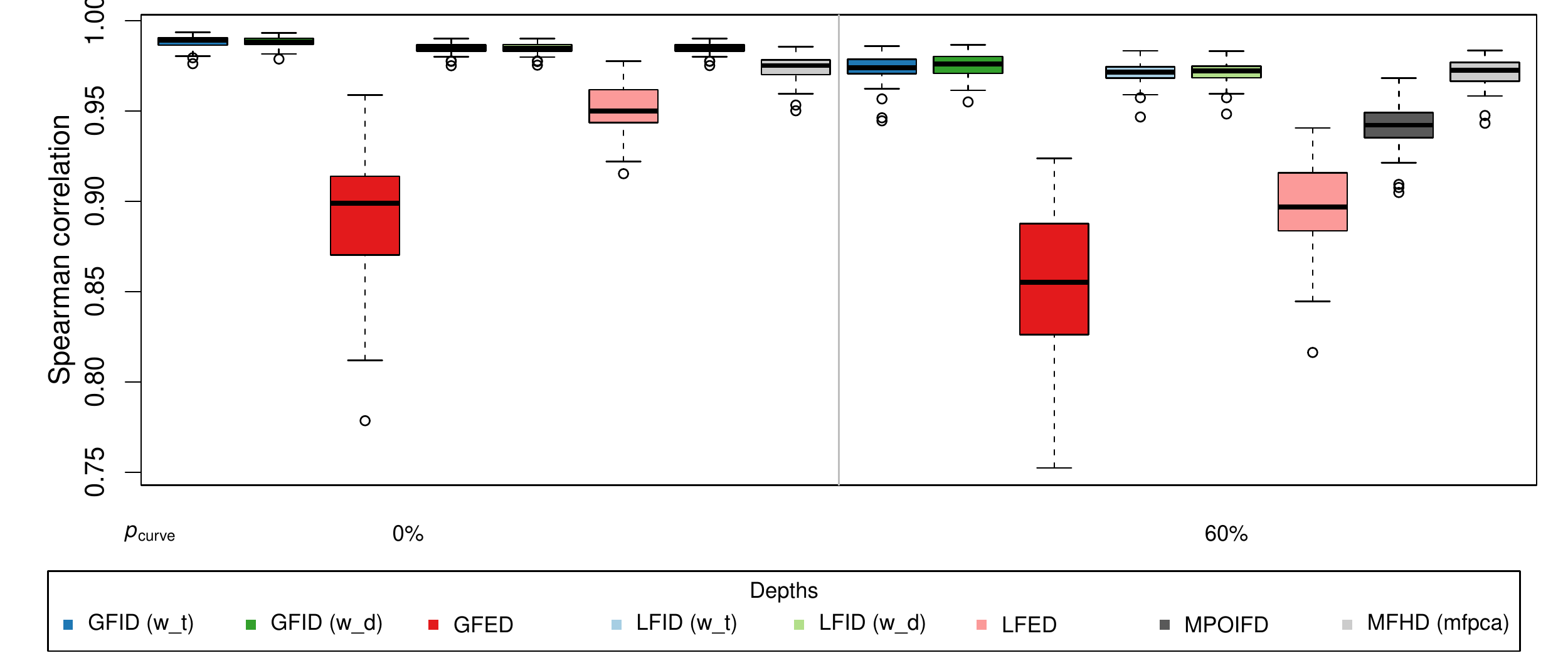}
\caption{Left column shows Model 1 with magnitude outlier I, and the right column shows Model 1 with shape outlier I. ASE of the estimated central curve (the first row), ASE of the 50\%-dispersion curve (the second row), the outlier proportion in the lowest 10\% depth region (the third row), and Spearman correlation (the fourth row). The sparseness type is point sparseness with a dense case ($p_{curve}=0\%$) and a high sparseness case ($p_{curve}\sim\mathcal{U}(0.4, 0.6)$).}
\label{m1}
\end{figure}

\begin{figure}[!b]
\centering
\includegraphics[width = 0.49\textwidth, height = 4cm]{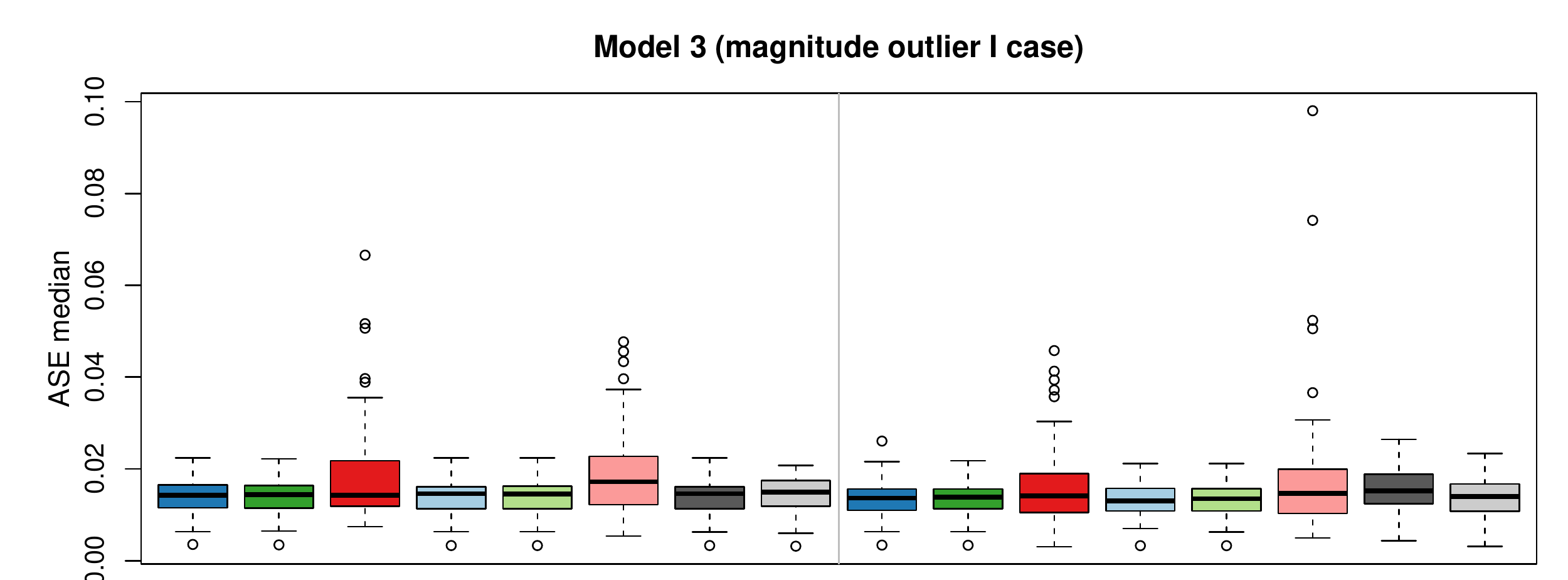}
\includegraphics[width = 0.49\textwidth, height = 4cm]{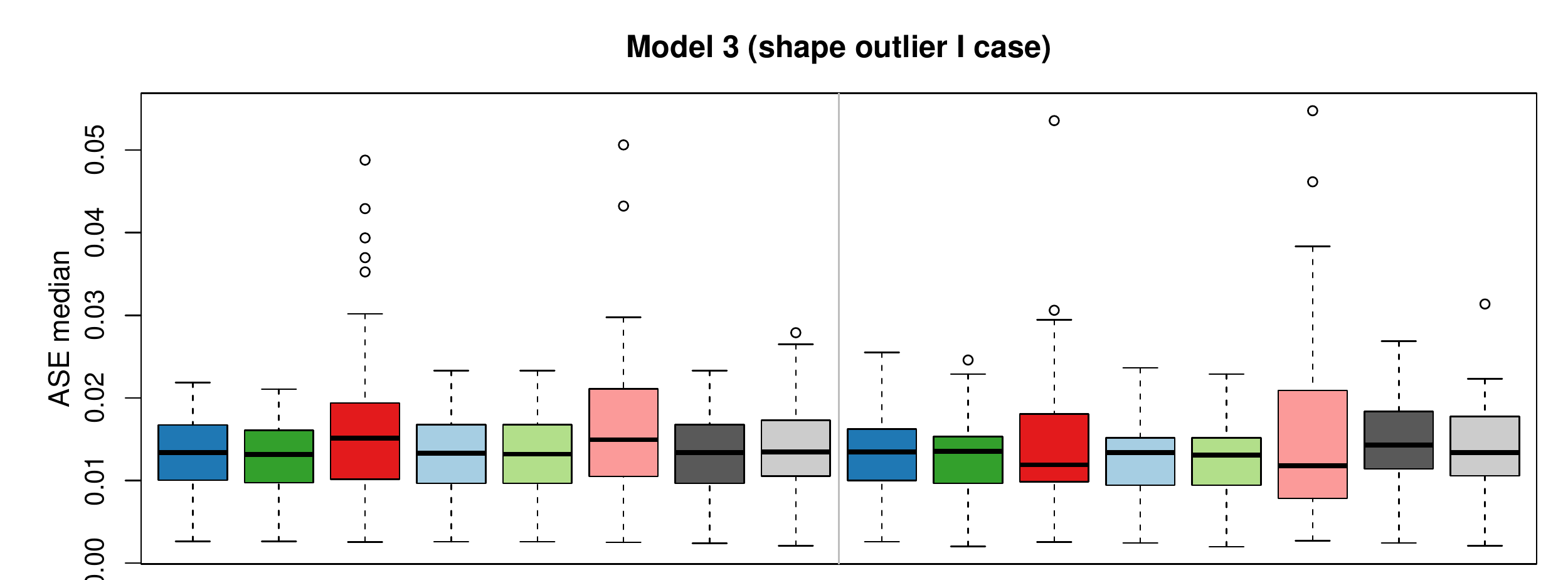}
\includegraphics[width = 0.49\textwidth, height = 3.5cm]{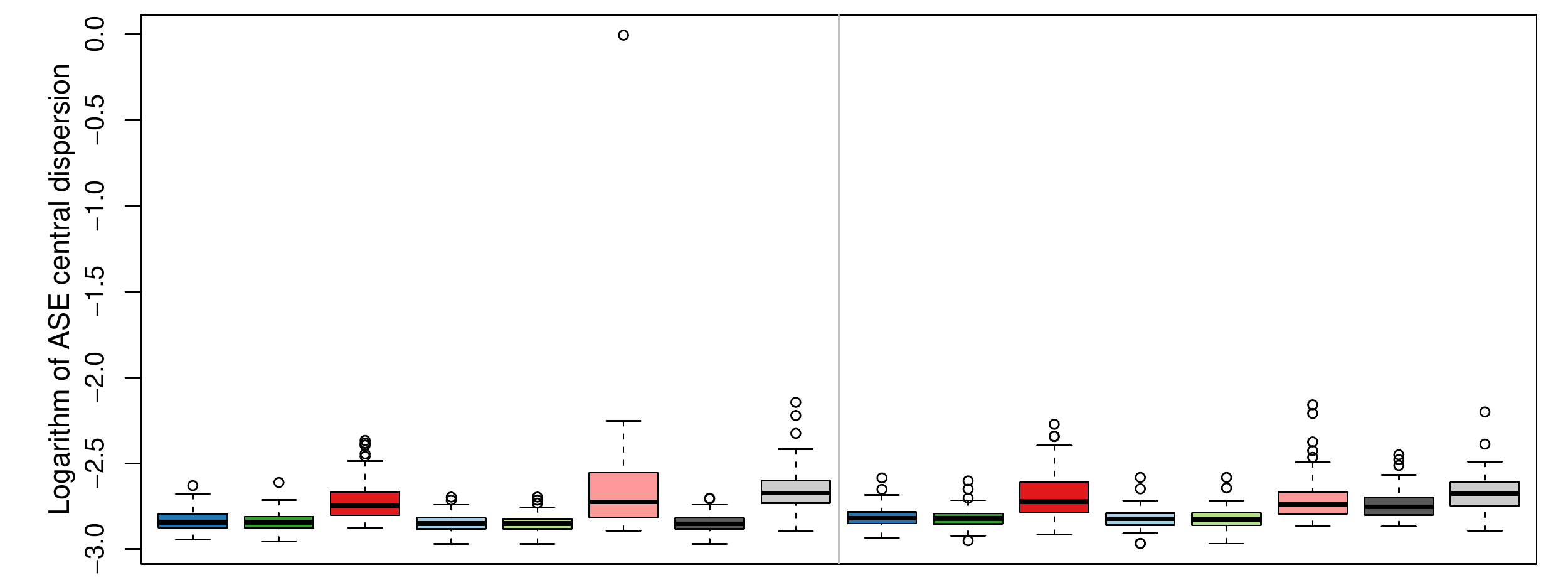}
\includegraphics[width = 0.49\textwidth, height = 3.5cm]{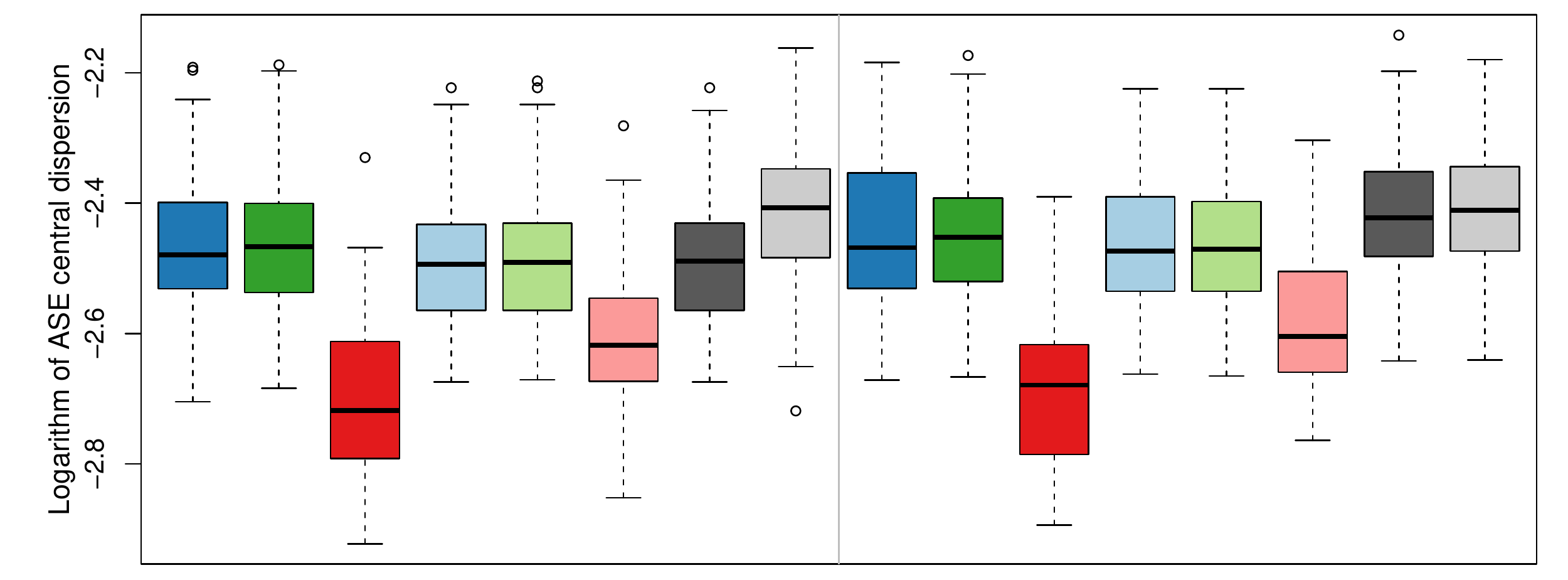}
\includegraphics[width = 0.49\textwidth, height = 3.5cm]{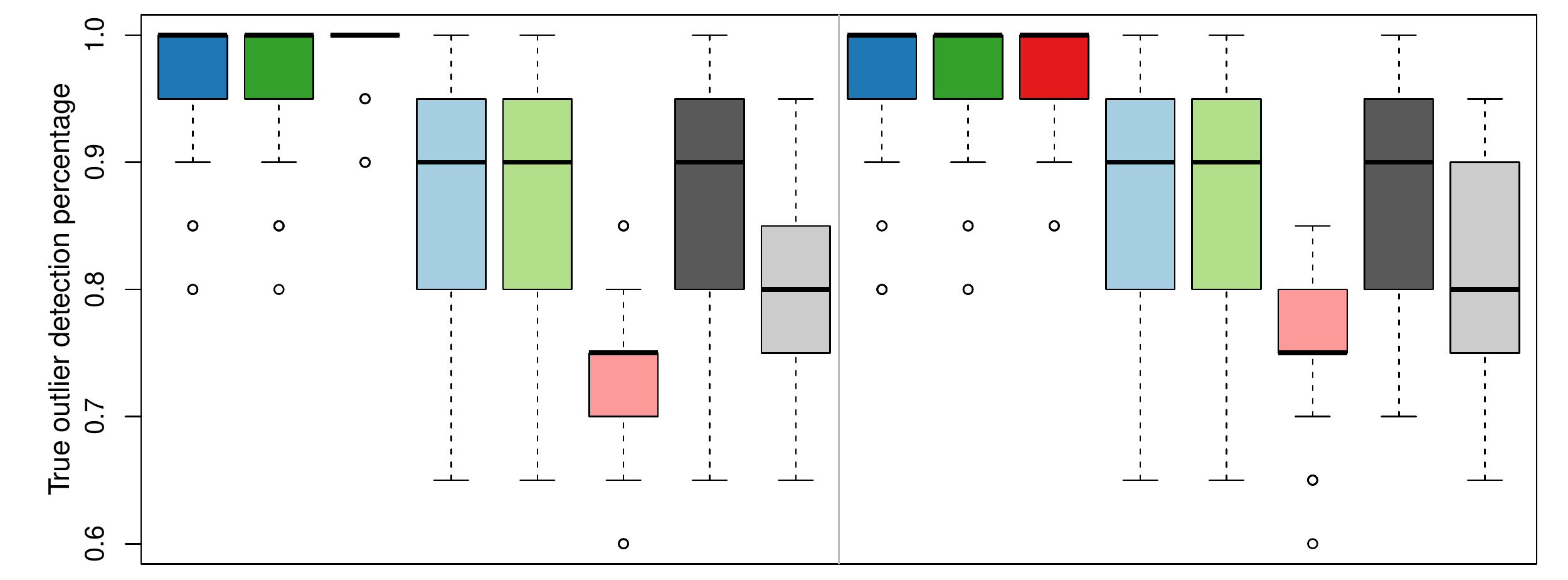}
\includegraphics[width = 0.49\textwidth, height = 3.5cm]{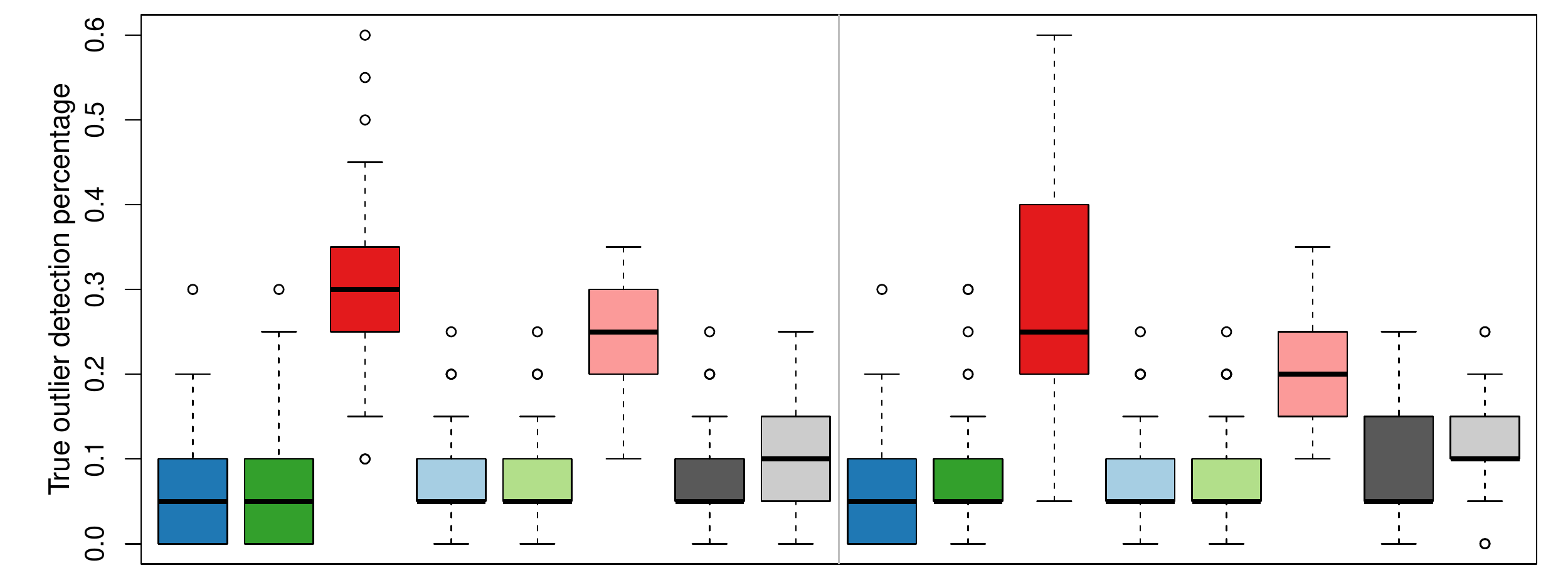}
\includegraphics[width = 0.49\textwidth, height = 4.2cm]{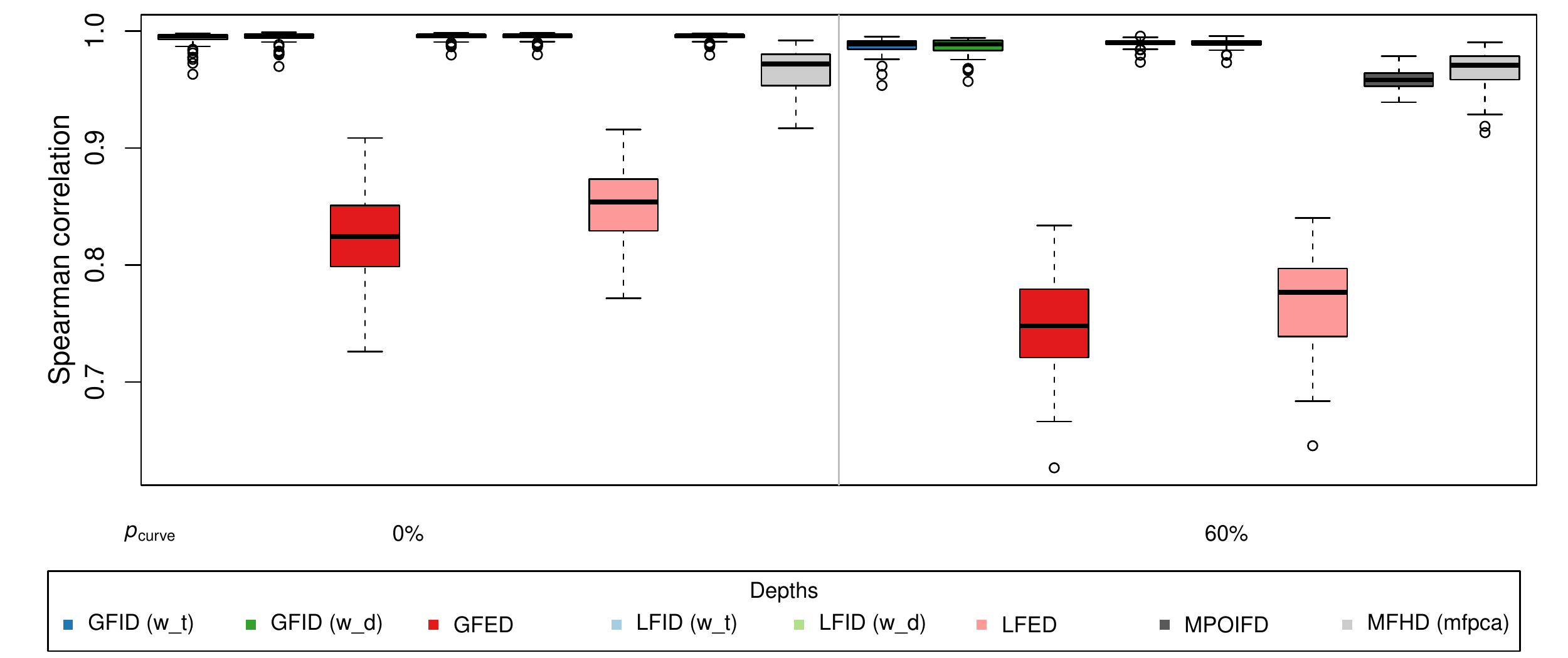}
\includegraphics[width = 0.49\textwidth, height = 4.2cm]{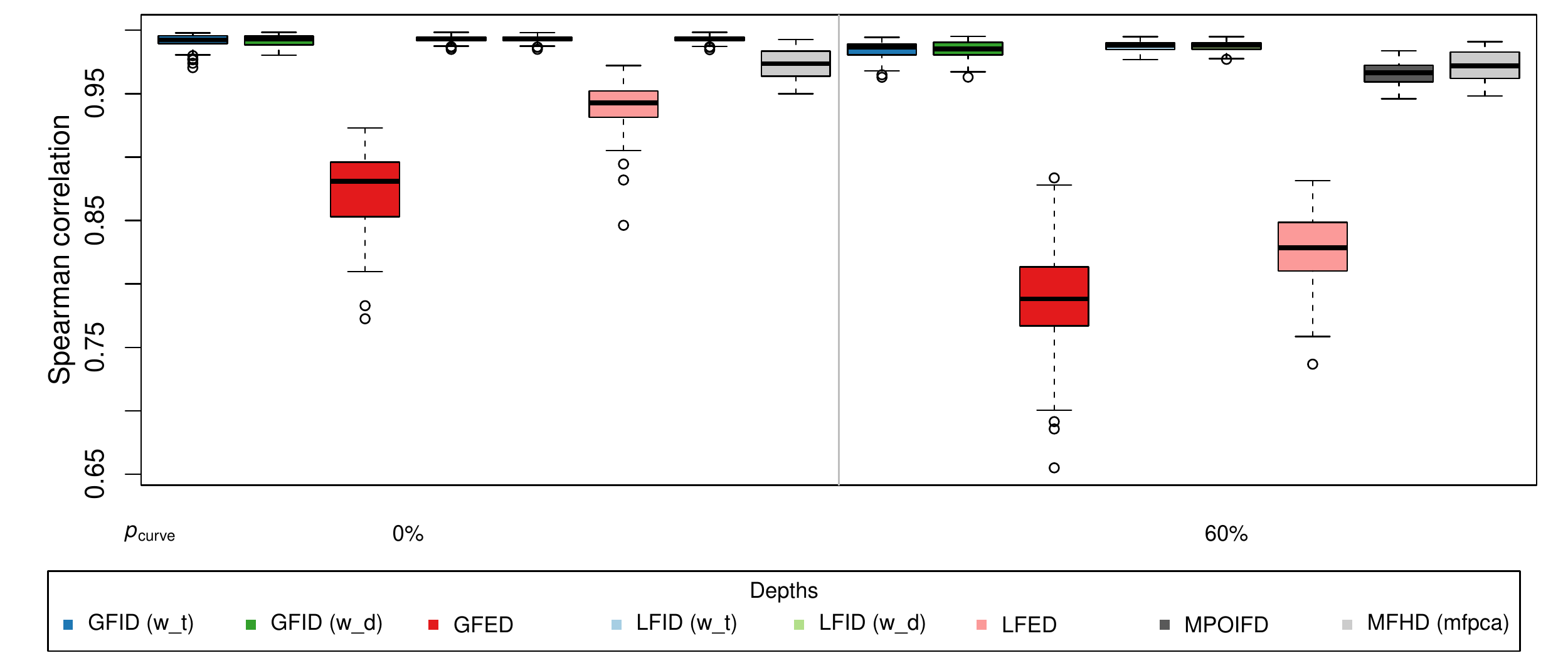}
\caption{Left column shows Model 3 with magnitude outlier I, and the right column shows Model 3 with shape outlier I. ASE of the estimated central curve (the first row), ASE of the 50\%-dispersion curve (the second row), outlier proportion in the lowest 10\% depth region (the third row), and Spearman correlation (the fourth row). The sparseness type is point sparseness with a dense case ($p_{curve}=0\%$) and a high sparseness case ($p_{curve}\sim\mathcal{U}(0.4, 0.6)$).}
\label{m3}
\end{figure}

We present the results of Models 1 and 3 under magnitude outlier I and shape outlier I with the point sparseness, in Figure \ref{simulation_visualization} for the simulation settings and Figures \ref{m1} and \ref{m3} for the assessment measures. The remaining results under different models, contaminations, and sparseness types are shown in the Supplementary Material. 

Among the GMFDs, the GMFID performs better than the GMFED in extracting the median (first row in Figures \ref{m1} and \ref{m3}), central region (second row in Figures \ref{m1} and \ref{m3}), as well as in maintaining the nonoutlier ranking association (fourth row in Figures \ref{m1} and \ref{m3}); in contrast, the GMFED usually is more sensitive to outliers than the GMFID and has better performance in outlier detection. 

MPOIFD (7th boxplot per case in Figures \ref{m1} and \ref{m3}) can only be applied when functional data are evaluated on a common time grid. Otherwise, MPOIFD cannot be applied directly and requires data preprocessing. The preprocessing can involve collecting individual time grids to form the union grids, separating union grids into several bins, and using binwise depths to approximate the pointwise depths. Under the setting that individual observation time points are not evaluated on common time grids, LMFID ($w_t$) is built exactly according to the aforementioned procedure; see the 4th boxplot per case in Figures \ref{m1} and \ref{m3}, and the estimation in Section \ref{estimate}. Hence, LMFID ($w_t$) can be regarded as the binwise MPOIFD.

We also notice that MFHD$_{mfpca}$ (the 8th boxplot per case in Figures \ref{m1} and \ref{m3}) performs well in extracting the median and central region and maintaining the ranking association, though it is not advantageous in outlier detection. Although the reconstructed data can describe the patterns of the nonoutliers, the shape of outliers usually cannot be captured correctly. Hence, the MFHD$_{mfpca}$ usually does not perform well in outlier detection. 

Compared to MPOIFD, GMFDs and LMFDs are applicable in more general designs of covariates. In terms of the ranking association of nonoutliers, GMFIDs and LMFIDs have advantages over MFHD$_{mfpca}$. Further, in terms of outlier detection, GMFED performs better than MPOIFD and MFHD$_{mfpca}$ in almost all cases. Notably, when GMFD does not show the best performance in outlier detection (see Model 1 (magnitude outlier I case), corresponding to the third row and the left panel of Figure \ref{m1}), GMFID is superior in outlier detection. Hence, GMFDs and LMFDs are proposed to validate the idea of defining depth notions for irregularly observed multivariate functional data based on original information. The proposed methods are not only alternatives to applying depths to reconstructed data defined on common grids but also perform better than the existing methods in most tasks.

We want to extend this discussion by considering different performances between GMFD and LMFD under each framework. Overall, GMFDs show better performance than LMFDs with increasing sparseness level. Usually, GMFID and LMFID have similar performances in extracting the median (the first row in Figures \ref{m1} and \ref{m3}), the central region (the second row in Figures \ref{m1} and \ref{m3}), and maintaining the ranking association (the fourth row in Figures \ref{m1} and \ref{m3}), whereas GMFED performs remarkably better than LMFED in central region extraction and outlier detection (the third row in Figures \ref{m1} and \ref{m3}). 

\begin{figure}[!b]
\centering
\includegraphics[width = 0.49\textwidth, height = 7cm]{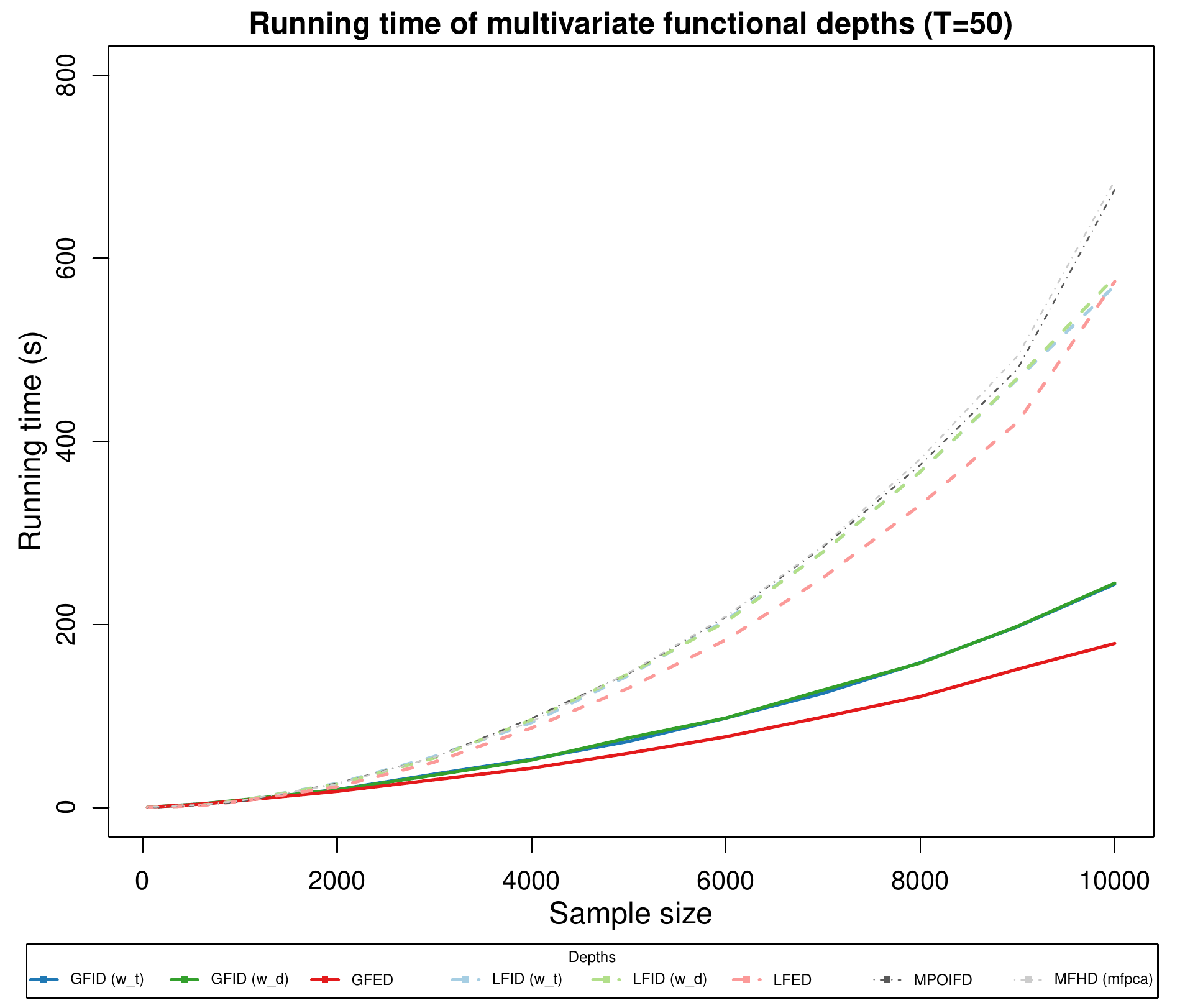}
\includegraphics[width = 0.49\textwidth, height = 7cm]{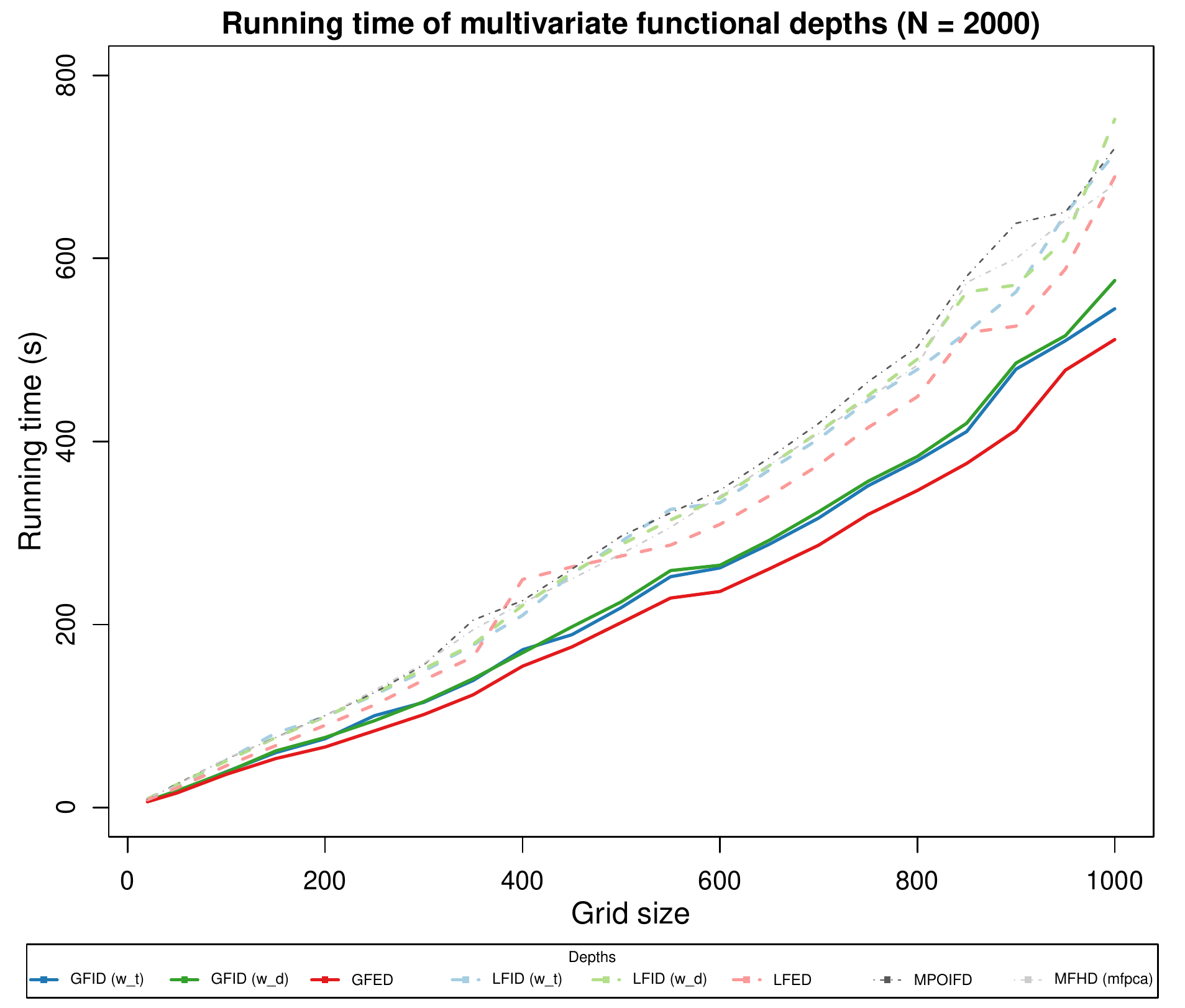}
\caption{Left figure shows the change in running time of multivariate functional depth calculation with the sample size when $T=50$ points are observed per subject. The second figure shows the change in running time of multivariate functional depth calculation with the grid size when $N=2000$ subjects are observed.}
\label{running_time}
\end{figure}
Although GMFDs and LMFDs are both potential alternatives to the previous MPOIFD and MFHD$_{mfpca}$, we recommend GMFDs because of their better performance in outlier detection and running time in big functional data. Figure \ref{running_time} shows the average running time of the aforementioned depths versus the change in sample size and grid size, i.e., $N$ and $T$, respectively. We see a remarkable difference in running time when the sample size is very large. Meanwhile, when the sample size is fixed at $N=2000$ and the design becomes denser, the GMFDs still perform better than the other methods. In this scenario, the time saved in calculating the multivariate depths of all scaled data is longer than the additional time required for scaling the binwise data for global MFDs. Comparatively, it takes much longer time to calculate multivariate depths under each small bin for local MFDs and for MPOIFD and MFHD$_{mfpca}$. 

Notably, the precision of extracting the median, central region, and ranking association does not change much when the average sparseness probability per subject increases from 0\% to 60\%. Thus, we have an alternative to the depth calculation for irregular time grids. To summarize our discussion, there are two main points to be emphasized. First, we can directly apply the GMFDs to irregularly observed multivariate functional data to explore the data ranking and outlier detection for such data type. Second, we can use GMFID (GMFED and GMFID) to estimate the central region (detect potential outliers). 

\section{\large \hspace{-.4cm}Visualization of Irregularly Observed Multivariate Functional Data}
\label{section4}
To demonstrate the effectiveness of the aforementioned depths in extracting the central region and detecting outliers, we improve the sparse functional boxplot (\citeauthor{qu2022sparse} \citeyear{qu2022sparse}) to a simplified version without data reconstruction. Two extreme cases need to be noted in advance. When the time grids for all subjects are common, the simplified sparse functional boxplot reduces to the original functional boxplot (\citeauthor{sun2011functional} \citeyear{sun2011functional}). When the time intervals for various subjects are purely disjoint, $\inf\limits_{t \in \mathcal{T}}g(t) > 0$ in Condition (B.1) is not satisfied. Then, the simplified sparse functional boxplot cannot be constructed. Similar to the functional boxplot, the following four characteristics are emphasized in the simplified visualization tools: the median, the central region, outliers if any, and nonoutlying bounds.
\subsection{Outlier Detection}
The (time) domain outliers $O_{\textrm{d}}$ are extracted before extracting the central tendency and flagging functional outliers. On the one hand, it is difficult to standardize rare observations at some time index because of insufficient size. On the other hand, domain outliers are usually neglected since the functional boxplot detects the outliers showing abnormality in responses rather than in domain, and other current detection tools assume that the curves belong to common time grids. Thus, domain outliers must be removed when considering multivariate curves versus time. Specifically, we check the boxplot of the time interval length per subject $I_i=\max(T_i)-\min(T_i)$~($i=1, \ldots, N$) and that of the logarithm of $I_i$, $\log I_i$ and label the outliers in the domain outliers $O_{\textrm{d}}$ if any.

In addition, a simple outlier detection method for irregularly observed multivariate functional data can be proposed according to the complementary outlier detection performances of GMFED and GMFID (see the first and third boxplots in Figures \ref{m1} and \ref{m3} in Section \ref{simulation_result}) when outliers exist. Assume the outliers do not exceed 10\% of all the samples. We find the curves that lie in the 10\% lowest GMFID region and GMFED region and flag their intersection as a set of the most potential outliers $O_{\textrm{p1}}$, and we label the remaining curves in their union as a set of the second-most potential outliers $O_{\textrm{p2}}$. Let the set of potential outliers $O_{\textrm{p}}$ be the union of $O_{\textrm{p1}}$ and $O_{\textrm{p2}}$. 

If we implement this potential outlier detection, then we let the set of outliers $O = O_{\textrm{d}} \cup O_{\textrm{p}}$. Otherwise, let $O=O_{\textrm{d}}$. Consider the new set $S'=S \setminus O$ with sample size $N'$. The 50\% central region and median are determined with respect to $S'$. Here, the 50\% central region $\bm{C}_{0.5}(t)$ for the $j$th ($j = 1,\ldots, p$) component evaluated at $t \in {\mathcal{T}}^{N}_{k}~(k=1, \ldots, \overline{T})$ is
$$C^{(j)}_{0.5}(t) = \{(t, X^{(j)}(t)): \min\limits_{r = 1, \ldots, \lceil N'/2 \rceil} Y^{(j)}_{[r]}(t) \leq X^{(j)}(t) \leq \max\limits_{r = 1, \ldots, \lceil N'/2 \rceil} Y^{(j)}_{[r]}(t), ~t \in \mathcal{T}^{N}_{k}\}$$ 
where $\lceil N'/2 \rceil$ is the smallest integer not less than $N'/2$, and $Y^{(j)}_{[r]}(t)$ is the $j$th component of the $r$th deepest curve of samples in $S'$ evaluated at $t$. Correspondingly, the upper bound of the central region evaluated at time $t \in {\mathcal{T}}^{N}_{k}$ is $Y^{(j)}_{ub, 0.5}(t) = \max\limits_{r = 1, \ldots, \lceil N'/2 \rceil} Y^{(j)}_{[r]}(t)$, and the lower bound evaluated at $t \in {\mathcal{T}}^{N}_{k}$ is $Y^{(j)}_{lb, 0.5}(t) = \min\limits_{r = 1, \ldots, \lceil N'/2 \rceil} Y^{(j)}_{[r]}(t)$.~We also define the range of $C^{(j)}_{0.5}$ at $t \in {\mathcal{T}}^{N}_{k}$ as the difference between the upper and lower bounds $R^{(j)}_{0.5}(t) := Y^{(j)}_{ub, 0.5}(t) - Y^{(j)}_{lb, 0.5}(t)$. The median $Y^{(j)}_{\lceil 1 \rceil}(t)$ is the curve with the highest depth in $S'$.

Let $O_{\textrm{f}}$ be the set of functional outliers. Then, a curve $\bm{Y}_{\textrm{o}}$ belongs to $O_{\textrm{f}}$ if the measurement of its $j$th component ${Y}_{\textrm{o}}^{(j)}$~($j=1, \ldots, p$) is higher than the summation of 1.5 times the range of $C^{(j)}_{0.5}$ and the upper bound of $C^{(j)}_{0.5}(t)$ (or lower than the difference between the lower bound of $C^{(j)}_{0.5}$ and 1.5 times the range of $C^{(j)}_{0.5}(t)$) at some time $t \in {\mathcal{T}}^{N}_{k}$. Then, $O=O \cup O_{\textrm{f}}$. The nonoutlying maximal (minimal) bound is established by connecting the maximal (minimal) points at all time indexes, excluding the outliers. That is, $Y^{(j)}_{ub}(t) = \max\limits_{\bm{Y} \in S \setminus O} Y^{(j)}(t)$, and $Y^{(j)}_{lb}(t) = \min\limits_{\bm{Y} \in S \setminus O} Y^{(j)}(t)$ for $t\in {\mathcal{T}}^{N}_{k}$ and $k=1, \ldots, \overline{T}$.

To summarize our operation, the implementation includes three stages. First, we remove the domain outliers $O_{\textrm{d}}$ if any. Second, we remove the potential outliers $O_{\textrm{p}}$. Finally, we apply the MFD to the remaining data respectively, determine the 50\% central region, and flag functional outliers $O_{\textrm{f}}$ according to the criterion of the central region range. By combining the simulation results given in Section \ref{section3} with GMFID, we can determine the median, central region, and functional outliers. In Section \ref{section5}, we present the cyclone tracks data, and illustrate the visualization tools with and without obtaining potential outliers~$O_{\textrm{p}}$.

\subsection{Simplified Sparse Functional Boxplot}
The simplified version of the sparse functional boxplot (\citeauthor{qu2022sparse} \citeyear{qu2022sparse}) underlines the sparseness probability within the central region $C^{(j)}_{0.5}$ without reconstructing sparse data. Since only the observed data are known, we calculate the observation counts in each small bin. We use $T_l$ equidistant points $t_{1}, \ldots, t_{T_l}$ to separate $\mathcal{T}^N$. For each $t \in [t_{w-1}, t_{w}]$~($w=2, \ldots, T_l$), we compute the observed proportion at each local bin $p_{o,w} = \frac{1}{N}\sum_{i=1}^N \sum_{l=1}^{T_i}\bm{1}\{t_{i,l} \in [t_{w-1}, t_{w}]\}$ assuming that $N$ samples are observed ideally. If $p_{o,w} >1$, then we replace $p_{o,w}$ with 1. Later, the proportion line of the $j$th component is denoted as $l^{(j)}(t, p_{o,w})=Y^{(j)}_{lb, 0.5}(t)+p_{o,w}R^{(j)}_{0.5}(t)$. The proportion line is smoothed and works as a separation in the central region. In the first row in Figure \ref{cyclone}, the lower region is colored in magenta, and the upper region is colored in grey. In addition, the 50\% proportion line $l^{(j)}(t, 0.5)$ is smoothed and labeled in dashed cyan in the central region.

\subsection{Simplified Intensity Sparse Functional Boxplot}
The original intensity sparse functional boxplot (\citeauthor{qu2022sparse} \citeyear{qu2022sparse}) shows the normalized sparseness intensity within the central region. Here, we use the observed intensity as a reflection of the sparseness intensity attributed to the lack of reconstructed sparse data. Assuming that we have altogether $K$ observed points data within $C^{(j)}_{0.5}$, we regard them as a spatial point pattern $\bm{z}^{(j)}_k:=\{(t_k, Y^{(j)}_k) \in C^{(j)}_{0.5}, k=1, \ldots, K\}$ with time $t_k$, and $Y^{(j)}_k$ as the corresponding observed point inside the central region. Then, the observed intensity at any point, $\bm{z}^{(j)} \in C^{(j)}_{0.5}$, can be expressed as $\lambda_o(\bm{z}^{(j)}) = e(\bm{z}^{(j)})\sum_{k=1}^{K} w_k \mathcal{K}(\bm{z}^{(j)}_k - \bm{z}^{(j)})$ from \citeauthor{diggle1985kernel} (\citeyear{diggle1985kernel}), where $\mathcal{K}$ is the Gaussian smoothing kernel, $e(\bm{z}^{(j)})$ is an edge correction factor, and $w_k$ are the weights. By default, the intensity values of the observed point patterns are expressed in estimated observed points per unit area. Then, $\lambda_o(\bm{z}^{(j)})$ is normalized such that $\overline{\lambda}_o(\bm{z}^{(j)}):= \lambda(\bm{z}^{(j)})/\max\limits_{k=1, \ldots, K}\{\lambda(\bm{z}^{(j)}_k)\}$. We infer the normalized sparseness intensity $\overline{\lambda}_{r}(\bm{z}^{(j)}) = 1 - \overline{\lambda}_o(\bm{z}^{(j)})$. In the second row in Figure \ref{cyclone}, the change in the color scale from magenta to white represents the change from the case of the least sparseness intensity ($\overline{\lambda}_{s}(\bm{z}^{(j)}) = 0$) to the case of the most sparseness intensity ($\overline{\lambda}_{s}(\bm{z}^{(j)}) = 1$).

\section{Application to Northwest Pacific Cyclone Tracks}
\label{section5}
\begin{figure}[b!]
\centering
\includegraphics[width = \textwidth, height = 6cm]{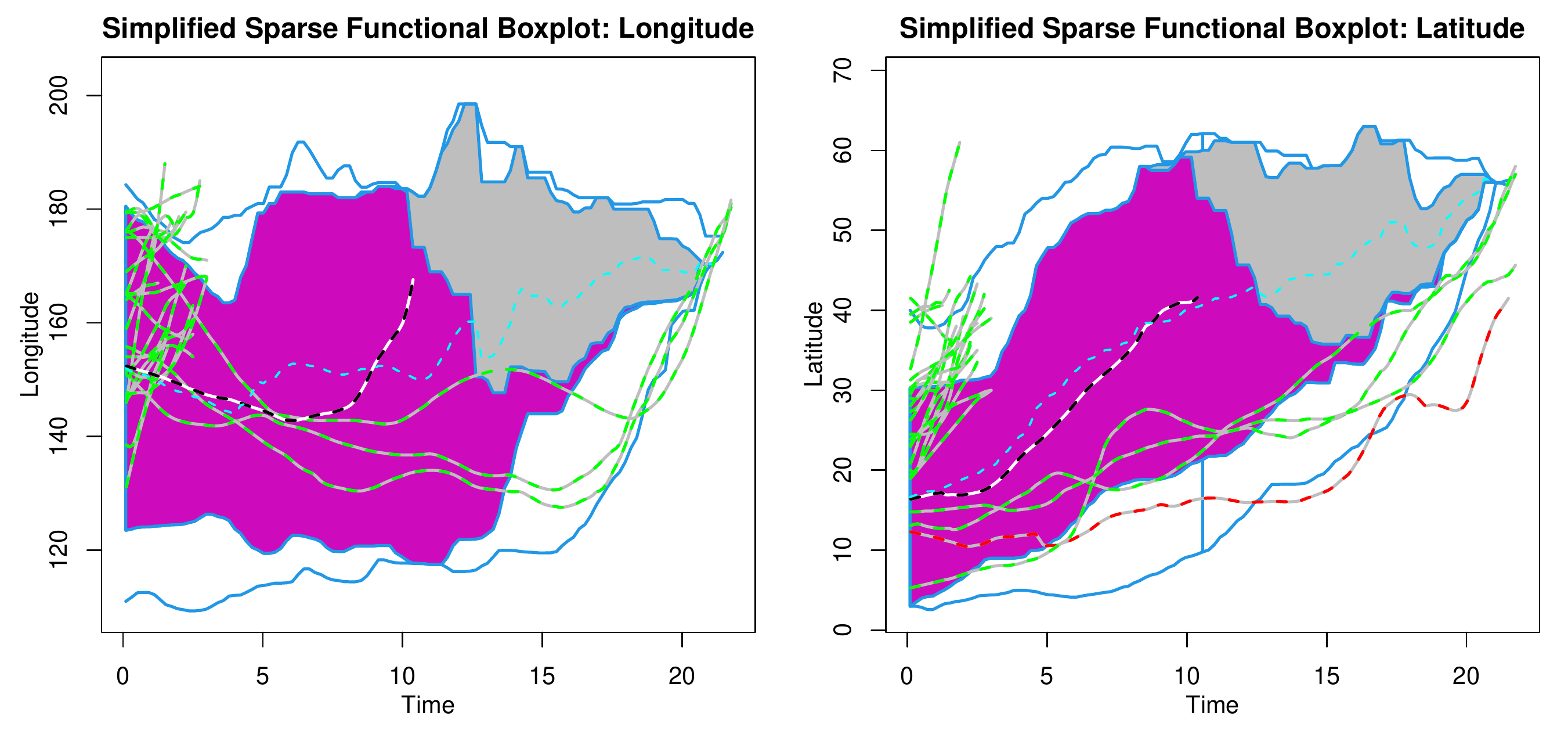}
\includegraphics[width = \textwidth, height = 6cm]{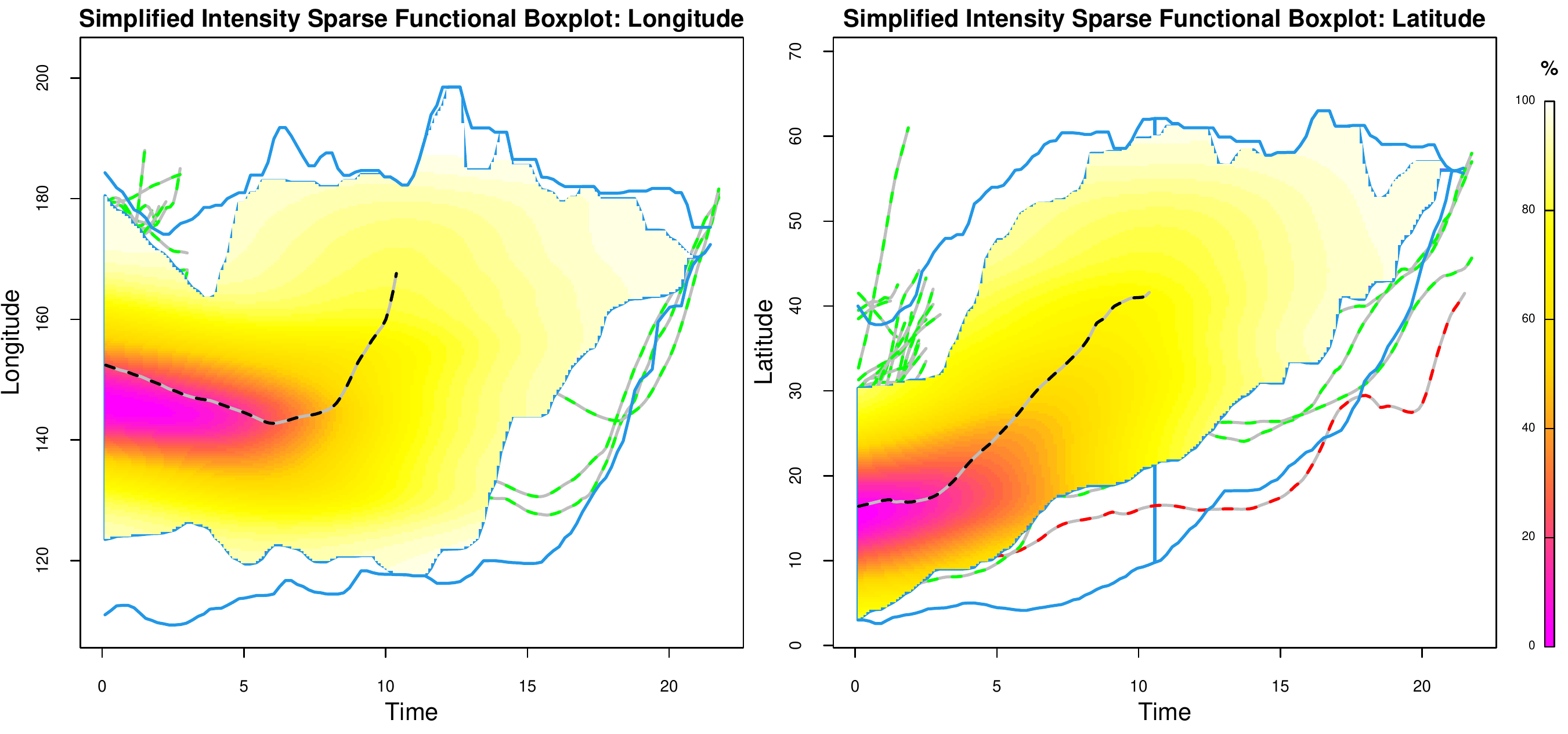}
\includegraphics[width = 0.49\textwidth, height = 6cm]{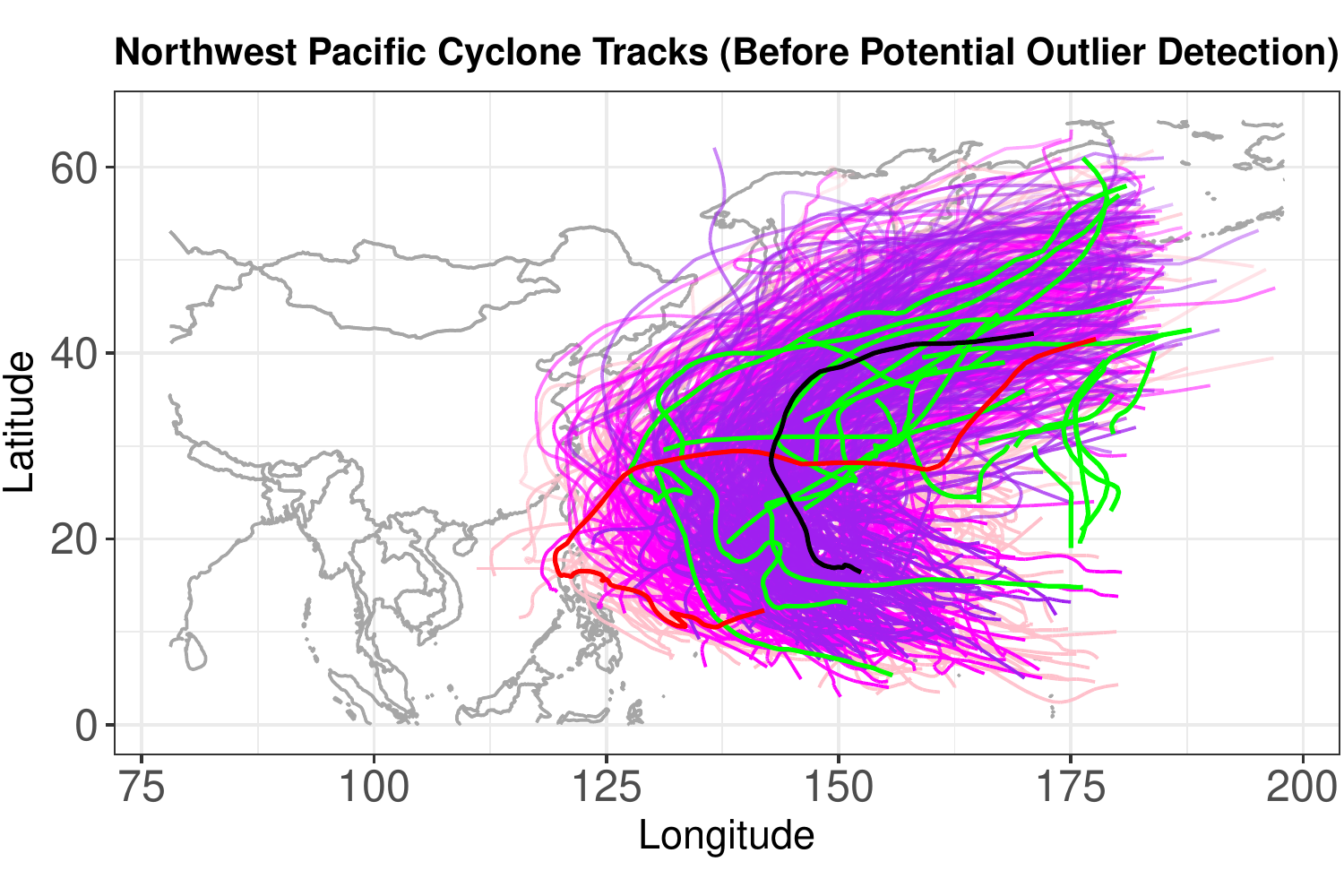}
\includegraphics[width = 0.49\textwidth, height = 6cm]{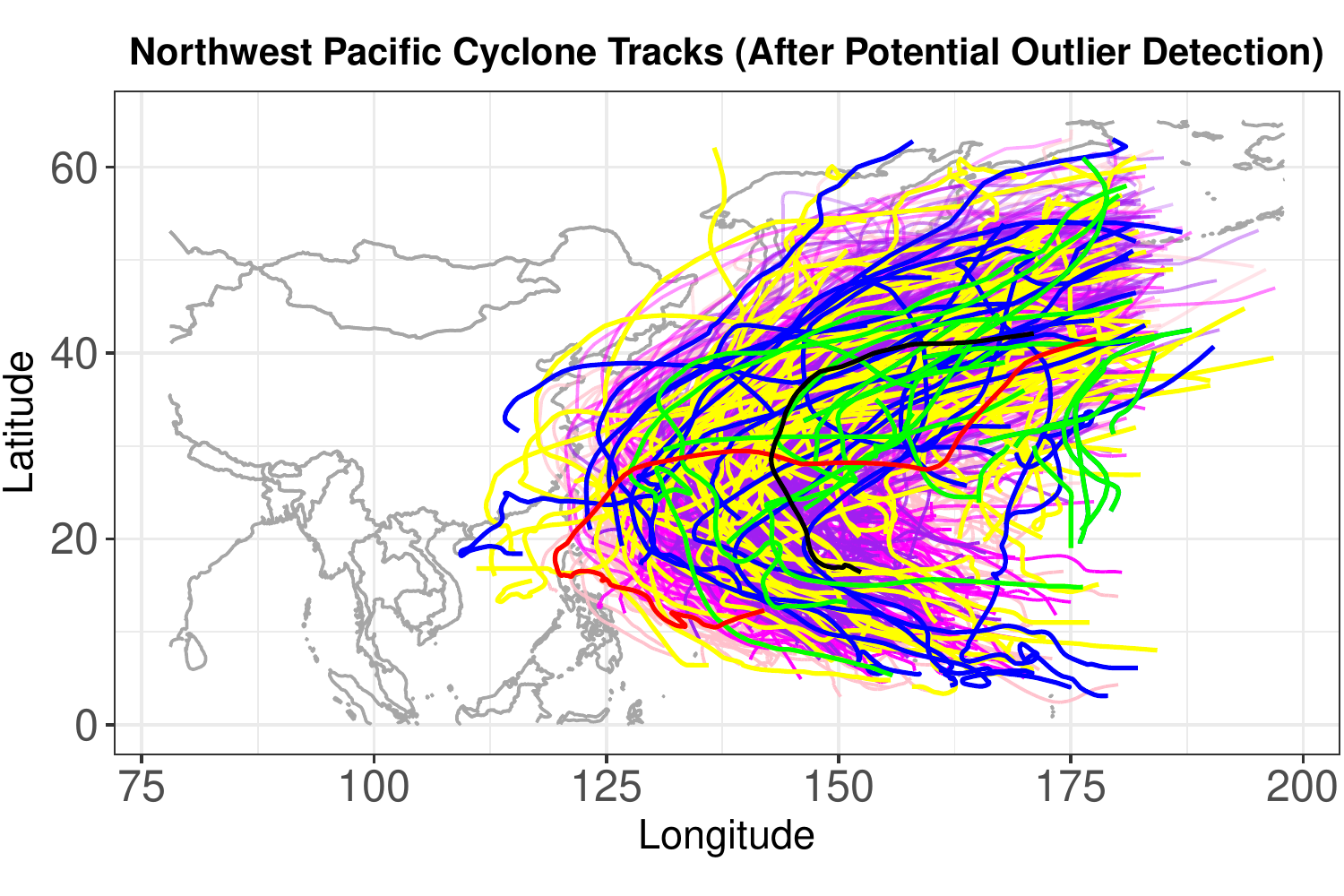}
\caption{The first row visualizes the components of the cyclone tracks data with the simplified sparse functional boxplot, and the second row visualizes the components of the tracks data with the simplified intensity sparse functional boxplot, while the third row shows the shape trajectory boxplot with the rank based on the global multivariate functional integrated depth (GMFID) before and after the potential outlier detection. The first, second, and third quartiles in the trajectory boxplot are colored in purple, magenta, and pink, respectively. The domain outliers (green), functional outliers (red), most potential outliers (blue), and the second-most potential outliers (yellow) are shown. The median is colored in black.}
\label{cyclone}
\end{figure}
We used a subset of the Northwest Pacific cyclone data (see Cluster 2 in the Northwest Pacific cyclone application in \citeauthor{qu2022robust} \citeyear{qu2022robust}); see Figure \ref{clusters}. There are 627 trajectories originating from the area close to the Philippines, passing by Japan, and ending in the Northern Pacific. The trajectories have a rotated U shape. In addition, the time interval per subject is variable, with the duration of cyclones lasting between 1 day to 22 days. Since each subject is under different time intervals and has various observations, there are challenges in the traditional multivariate functional depth application. We aim to use GMFID to rank the trajectories, extract the median and central tendency, and flag possible outliers. Additionally, we use the combination of GMFED and GMFID to find potential outliers within the 10\% most extreme curves from any of the GMFDs.

Before applying GMFID, 23 domain outliers are flagged, with three cyclones lasting more than 21 days and the remaining cyclones lasting less than three days. Their shapes show deviation from the U shape (third row, left panel in Figure \ref{cyclone}). Then, from the first row in Figure \ref{cyclone}, i.e., the simplified sparse functional boxplots of the locations, it can be seen that when the time does not exceed 10 days, sufficient observations are made in each small local bin. However, when the time ranges from 10$-$15 days, the observed probability decreases significantly from 100\% to around 20\%. From the second row in Figure \ref{cyclone}, i.e., the simplified intensity sparse functional boxplots of the locations, the most observed points are concentrated for time less than 10 days, the longitude between 140$\degree$ and 155$\degree$, and the latitude between 10$\degree$ and 23$\degree$. 

The GMFID selects a median lasting around 10 days as a representative pattern. Among the first 75\% quantile region in our version of the trajectory boxplot (the third row, left panel in Figure \ref{cyclone}, \citeauthor{yao2020trajectory} \citeyear{yao2020trajectory}), we see that the first 25\% central region barely covers any land, and the 25\% to 75\% quantile region covers most island areas in Eastern Asia, such as the Philippines, Taiwan, Japan, and the Korean Peninsula. Only one functional outlier is detected from the simplified sparse functional boxplot (the first row, right panel in Figure \ref{cyclone}) because of its magnitude abnormality in the latitude. This outlier lasts more than 21 days, and because of the slow movement in terms of the latitude, the cyclone track differs from the U-shape pattern (red curve in the third row, left panel in Figure \ref{cyclone}). 

From the simulation study, the combination of GMFID and GMFED may flag potential magnitude and shape outliers if we focus on the 10\% most extreme curves from both depths. The adapted trajectory boxplot obtained after potential outlier detection is shown in the third row, the right panel in Figure \ref{cyclone}. Altogether, there are 32 most potential outliers (colored in blue) and 62 second-most potential outliers (colored in yellow) detected, together with previous domain outliers (colored in green) and one functional outlier (colored in red). If there are intersections between the potential outliers and the domain/functional outliers, then we use the color of domain/functional outliers to cover the color of potential outliers. We see that many shape outliers are detected. These outliers may be cyclones originating far from the Philippines, those ending somewhere far from the Northern Pacific, or those revealing patterns different from the rotated U shape like the median. Although we see numerous shape outliers, the need for further improvement in depth-based shape outlier detection with solid theoretical/numerical support is expected.
\vspace{-.2cm}

\section{Discussion}
\label{section6}
\vspace{-.4cm}
In this study, we examined two different frameworks for multivariate functional depths based on multivariate depths: the first one is the integrated depth, and the second is the extremal depth. In each framework, local and global multivariate functional depths can be specified according to whether the multivariate depths are computed from the original data or the normalized data. The term local MFD is coined to show contrast with the global MFD. In GMFDs, the multivariate depths can be computed only once by computing the depth of any normalized data with respect to all the normalized data across time indexes. However, in LMFDs, the multivariate depths need to be calculated at each fixed time index (i.e., the observations should be collected at some fixed time, the depth of any data should be calculated with respect to such observations, and the above procedures should be implemented for all possible time indexes).

Under the finite sample cases, each sample may be defined with different time intervals and observation locations. This poses applicational challenges in obtaining the multivariate depths in MFD estimation. Different strategies are possible when obtaining multivariate depths in GMFDs and LMFDs. First, multivariate depths in GMFDs can be well estimated as long as the bin-based sample mean and sample covariance are good estimates of the population mean and covariance. Second, multivariate depths in LMFDs replace the estimation of pointwise depths at each time index with the estimation of binwise depths at each small bin. Thus, suitable bin construction is key for estimating multivariate depths for GMFDs and LMFDs.

To ensure a robust depth under outlier contamination and flexible time grids, we recommend combining GMFID and GMFED. The advantage of GMFID is that it can extract the median and central region well and can maintain the rank association of nonoutliers, whereas GMFED is advantageous in outlier detection. In addition, GMFDs can be calculated within a shorter time than LMFDs and the current MFHD based on the reconstructed data. Furthermore, GMFDs can be applied to time grids that are more general compared to the other comparative depth, i.e., MPOIFD. 

The simplified sparse functional boxplot and simplified intensity sparse functional boxplot are provided to visualize irregularly observed multivariate functional data. The simplified sparse functional boxplot shows the binwise observed probability in the central region, and the simplified intensity sparse functional boxplot shows the normalized sparseness intensity in the central region. The domain outliers and functional outliers can be detected using such visualization tools. Moreover, potential outliers can be flagged from the 10\% most extreme curves obtained from GMFID and GMFED.

There are several directions for future research. The first direction to explore is depth-based shape outlier detection method for irregularly observed multivariate functional data. We can see that the current method to detect potential outliers may include nonoutliers, and hence, a data-driven way to determine the outlier threshold in the samples is needed. Another direction is the exploration of other MFD frameworks, and the theoretical and numerical comparison between its global and local formulas. The $L^p$ depth or $L^\infty$ depth (\citeauthor{long2015study} \citeyear{long2015study}) may be another good building block as the multivariate depth.
\bigskip
\begin{center}
{\large\bf SUPPLEMENTARY MATERIAL}
\end{center}

{\bf Title:} Supplementary Material (proofs): Global Depths for Irregularly Observed Multivariate Functional Data (.pdf). {\bf R-code for global depths:} Depthnotion contains code to perform the ranking methods described in the article. The package also contains the cyclone data used as an example in the article (Depthnotion zipped tar file). {\bf Cyclone data set:} Data set used in the illustration of the global multivariate functional depths in Section 5 (.RData file).

\setlength{\bibsep}{6.7pt}
\bibliographystyle{apalike2}
\bibliography{ref} 
\end{document}